\documentstyle[12pt,psfig]{article}
\def \bt{\bar T}
\def \bc{\bar C}
\def \be{\bar E}
\def \ba{\bar A}
\def \tt{\tilde T}
\def \tc{\tilde C}
\def \te{\tilde E}
\def \ta{\tilde A}
\def\that{{\hat T}}
\def\chat{{\hat C}}
\def\phat{{\hat P}}
\def\ehat{{\hat E}}
\def \cn{Collaboration}

\newcommand{\thspace}{\kern.08333em}

\def \beq{\begin{equation}}
\def \eeq{\end{equation}}
\def \beqn{\begin{eqnarray}}
\def \eeqn{\end{eqnarray}}

\def \s{\sqrt{2}}

\def \v#1#2{V_{#1#2}}

\begin{document}
\rightline{TECHNION-PH-95-10}
\rightline{UdeM-GPP-TH-95-24}
\rightline{EFI-95-09}
\rightline{hep-ph/9504326}
\rightline{March 1995}
\bigskip
\bigskip
\centerline{\bf BROKEN SU(3) SYMMETRY IN TWO-BODY $B$ DECAYS}
\bigskip
\centerline{\it Michael Gronau}
\centerline{\it Department of Physics}
\centerline{\it Technion -- Israel Institute of Technology, Haifa 32000,
Israel}
\medskip
\centerline{and}
\medskip
\centerline{\it Oscar F. Hern\'andez\footnote{e-mail:
oscarh@lps.umontreal.ca} {\rm and} David London\footnote{e-mail:
london@lps.umontreal.ca}}
\centerline{\it Laboratoire de Physique Nucl\'eaire}
\centerline{\it Universit\'e de Montr\'eal, Montr\'eal, PQ, Canada H3C 3J7}
\medskip
\centerline{and}
\medskip
\centerline{\it Jonathan L. Rosner}
\centerline{\it Enrico Fermi Institute and Department of Physics}
\centerline{\it University of Chicago, Chicago, IL 60637}
\bigskip
\centerline{\bf ABSTRACT}
\medskip
\begin{quote}
The decays of $B$ mesons to two-body hadronic final states are analyzed
within the context of broken flavor SU(3) symmetry, extending a previous
analysis involving pairs of light pseudoscalars to decays involving one or
two charmed quarks in the final state. A systematic program is described
for learning information {}from decay rates regarding (i) SU(3)-violating
contributions, (ii) the magnitude of exchange and annihilation diagrams
(effects involving the spectator quark), and (iii) strong final-state
interactions. The implication of SU(3)-breaking effects for the extraction
of weak phases is also examined. The present status of data on these
questions is reviewed and suggestions for further experimental study are
made.
\end{quote}
\newpage

\centerline{\bf I. INTRODUCTION}
\bigskip

Recently \cite{BPP} we analyzed the decays of $B$ mesons to two light
pseudoscalar mesons $P$ within the context of flavor SU(3)
\cite{DZ,SW,Chau}. We proposed that information on phases of the
Cabibbo-Kobayashi-Maskawa (CKM) matrix could be obtained {}from the study
of time-independent measurements of decay rates, and found that the SU(3)
relations were of use in interpreting and anticipating
CP-violating asymmetries in these decays.

The analyses in \cite{BPP} made use of an overcomplete graphical
description of amplitudes involving dominant tree $T$, color suppressed
$C$, and penguin $P$ contributions, and smaller exchange $E$, annihilation
$A$, and penguin annihilation $PA$ terms. Particularly useful relations
followed {}from the neglect of these last three terms. For a $B$ meson to
decay via these diagrams directly the two quarks in the meson must find
each other, and hence the contributions of these diagrams were expected to
be suppressed by a factor of $f_B/m_B \approx 5\%$. Tests of this
assumption that relied on $B$ decays to the pseudoscalar mesons were
proposed in \cite{BPP}.

One can also test for the absence of exchange and annihilation graphs in
the decays ~\cite{DZ,SW,Chau} of $B$'s to one light pseudocalar $P$ and one
charmed meson $D$. (In these processes there is no analogue of the penguin
annihilation graph.) Furthermore, various SU(3)-breaking effects can be
studied in a manner not possible when both final-state mesons are light.
Since a single product of CKM elements is involved in such decays, relative
phases between amplitudes are a signal of final-state interactions, which
thus may be probed with the help of amplitude triangles \cite{HY}. When two
charmed quarks occur in the final state, as in the decays $B \to D \bar D$
or $B \to \eta_c P$, the analysis becomes even simpler.

The strangeness-preserving processes $B \to P \bar D$, involving the CKM
matrix element product $V_{cb}^* V_{ud}$, have typical branching ratios of
several parts in $10^3$. They dominate the much rarer $B \to PP$ processes,
which involve $V_{ub}$ and are expected to have branching ratios of order
$10^{-5}$. The strangeness-changing processes $B \to P \bar D$, involving
the combination $V_{cb}^* V_{us}$, as well as the rarer processes $B \to P
D$, involving the combinations $V_{ub}^* V_{cs}$ or $V_{ub}^* V_{cd}$, also
provide useful information, as do the decays of $B$ mesons to $D \bar D$ or
$\eta_c P$ final states.

Several issues arose in \cite{BPP} which can be addressed in part by
extending the analysis to decays involving one or two charmed quarks in the
final state. We address these issues in the present paper:

(1) How large are SU(3)-breaking effects in two-body $B$ meson decays?

(2) Are contributions due to exchange ($E$) and annihilation ($A$) diagrams
really negligible?

(3) Can one determine final-state interactions in a manner independent of
CKM phases?  One such determination involves the decays $B^+ \to \pi^+ \bar
D^0$, $B^0 \to \pi^+ D^-$, and $B^0 \to \pi^0 \bar D^0$ \cite{HY}.

In Refs.~\cite{PRL,PLB} we discussed several ways in which weak CKM phases
could be determined using SU(3) triangle relations involving a variety of
$B\to PP$ processes. In this paper we discuss how these analyses are
affected by SU(3)-breaking effects. We will see that, for the most part,
SU(3) breaking can be taken into account in a systematic way. In
Ref.~\cite{DH}, the question was raised as to the importance of electroweak
penguin diagrams in the determination of weak phases. Although this is an
important point, it is somewhat orthogonal to the main thrust of the
present work. We therefore discuss it in a separate paper \cite{EWP}.

The impatient reader may turn directly to our conclusions (Sec.~VIII) for
the answers (many of which will require new measurements) to the above
questions. For more leisurely perusal, the following sections may be of
interest as well.

In Section II we review our SU(3) analysis \cite{BPP} of $B \to P P$
processes, and extend it to $B \to P \bar D$, $B \to P D$, $B \to D \bar
D$, and $B \to\eta_c P$ decays. The SU(3) analysis will lead to many useful
relations. For all except the $B \to PP$ processes, equivalent relations
can be obtained by simply replacing one or both of the pseudoscalar mesons
in the final state with a vector meson. Of course, when both are vector
mesons, amplitude relations will hold separately for different helicity or
angular momentum states, limiting their usefulness.

The language employed involves a graphical notation equivalent to
decomposition into SU(3) representations. We introduce this notation and
apply it to the case of first-order SU(3) breaking in Sec.~III.
Measurements which test these relations, both in the presence and in the
absence of exchange ($E$) and annihilation $(A$) contributions, are noted
in Sec.~IV. In Sec.~V we examine how SU(3)-breaking effects affect the
extraction of weak CKM phases. We discuss amplitude triangle relations and
their implications for strong final-state interactions in Sec.~VI. The
present status of relevant data on two-body $B$ decays, and some future
experimental prospects, are reviewed in Sec.~VII.

In our approach, the graphical description is used to implement flavor
SU(3) symmetry and linear SU(3) breaking in the most general form. Some of
our relations follow purely from this linearly broken symmetry, while
others depend on an additional (testable) dynamical assumption that permits
us to ignore certain contributions. This is complementary to the
model-dependent studies of two-body $B$ decays carried out in the past
\cite{twobodyB}. Such model calculations are based on further assumptions
of factorizable hadronic matrix elements of the effective Hamiltonian and
on specific hadronic form factors. This leads to stronger predictions than
in our approach -- absolute branching ratios, for example. However, the
model-dependent description is also expected to involve a number of
different kinds of uncertainties \cite{uncertainties}, so that it is
probably only sufficient for order-of-magnitude rate estimates.

\bigskip
\centerline{\bf II. NOTATION AND SU(3) DECOMPOSITION}
\bigskip

\leftline{\bf A. Definitions of states}
\bigskip

We recapitulate some results of [1]. Taking the $u,~d$, and $s$ quarks to
transform as a triplet of flavor SU(3), and the $-\bar u,~\bar d$, and
$\bar s$ to transform as an antitriplet, we define mesons in such a way as
to form isospin multiplets without extra signs:
\beq
\pi^+ \equiv u \bar d~~,~~~\pi^0 \equiv (d \bar d - u \bar u)/\s~~,~~~
\pi^- \equiv - d \bar u~~~,
\eeq
\beq
K^+ \equiv u \bar s~~,~~~K^0 \equiv d \bar s~~~,
\eeq
\beq
\bar K^0 \equiv s \bar d~~,~~~K^- \equiv - s \bar u~~~.
\eeq

For reasons discussed in more detail in [1], we do not consider decays
involving $\eta$ or $\eta'$ in the present paper. Since these states are
octet-singlet mixtures, we would have to introduce additional SU(3) reduced
matrix elements or additional graphs to describe such decays.

The $B$ mesons and their charge-conjugates are defined as
\beqn
B^+ \equiv \bar b u~~~,~~~B^0 &\equiv& \bar b d~~~,~~~ B_s \equiv \bar b s
{}~~~, \nonumber \\
B^- \equiv -b \bar u~~~,~~~\bar B^0 &\equiv& b \bar d~~~,~~~\bar B_s \equiv
b \bar s ~~~.
\eeqn
Charmed mesons are taken to be
\beqn
D^0 \equiv - c \bar u~~~,~~~D^+ &\equiv& c \bar d~~~,~~~D_s^+ \equiv
c \bar s~~~, \nonumber \\
\bar D^0 \equiv \bar c u~~~,~~~D^- &\equiv& \bar c d~~~,~~~ D_s^- \equiv
\bar c s~~~.
\eeqn

\bigskip

\leftline{\bf B. Decomposition in terms of SU(3) amplitudes}
\bigskip

{\it 1. $B \to PP$ decays} were discussed in \cite{BPP}. The weak
Hamiltonian operators associated with the transitions  $\bar b \to \bar u u
\bar q$ and $\bar b \to \bar q$ ($q = d$ or $s$) transform as a ${\bf
3^*}$, ${\bf 6}$, or ${\bf 15^*}$ of SU(3). When combined with the triplet
light quark in the $B$ meson, these operators then lead to the following
representations in the direct channel:
\beq
{\bf 3^*} \times {\bf 3} = {\bf 1} + {\bf 8}_1~~~,
\eeq
\beq
{\bf 6} \times {\bf 3} = {\bf 8}_2 + {\bf 10}~~~,
\eeq
\beq
{\bf 15^*} \times {\bf 3} = {\bf 8}_3 + {\bf 10^*} +{\bf 27}~~~.
\eeq

These representations couple to the symmetric product of two octets (the
pseudoscalar mesons), containing unique singlet, octet, and ${\bf 27}$-plet
representations, so that the decays are characterized by one singlet, three
octets, and one ${\bf 27}$-plet amplitude. Separate amplitudes apply to the
cases of strangeness-preserving and strangeness-changing transitions.

{\it 2. $B \to P \bar D$ decays}, involving the quark subprocess $\bar b
\to \bar c u \bar q~(q = d$ or $s$), are characterized by a weak
Hamiltonian transforming as a flavor octet. When combined with the initial
light quark ({\bf 3}), this leads to final states transforming as {\bf 3},
${\bf 6^*}$, and {\bf 15} representations of SU(3). These are also the
representations formed by the combination of the final octet light
pseudoscalar meson and triplet $\bar D$ meson. Thus, there are three
independent SU(3) amplitudes, transforming as {\bf 3}, ${\bf 6^*}$, and
{\bf 15}, for these decays.

{\it 3. $B \to P D$ decays}, involving the quark subprocess $\bar b \to
\bar u c \bar q~(q = d$ or $s$), are characterized by a weak Hamiltonian
transforming as a {\bf 3} or ${\bf 6^*}$ representation. When combined
with the initial light quark ({\bf 3}), this leads to the following
representations:
\beq
{\bf 3} \times {\bf 3} = {\bf 3}^{\bf *}_1 + {\bf 6}~~~,
\eeq
\beq
{\bf 6^*} \times {\bf 3} = {\bf 3}^{\bf *}_2 + {\bf 15^*}~~~,
\eeq
which each have unique couplings to the final light pseudoscalar ({\bf 8})
and charmed meson (${\bf 3^*}$), whose tensor product involves ${\bf 3^*}$,
${\bf 6}$, and ${\bf 15^*}$ representations. Thus, these processes are
characterized by four invariant amplitudes.

{\it 4. $B \to D \bar D$ and $B \to \eta_c P$ decays} are characterized by
transitions giving rise to a single light antiquark, transforming as an
antitriplet. When combined with the initial quark, this antiquark can form
a singlet or an octet in the direct channel. Thus, there will be two
SU(3)-invariant amplitudes ({\bf 1} + {\bf 8}) characterizing the decays $B
\to D \bar D$ but only one ({\bf 8}) characterizing the decays $B \to
\eta_c P$, where $P$ is an octet member.

\bigskip
\leftline{\bf C. Decomposition in terms of diagrams}
\bigskip

\begin{figure}
\centerline{\psfig{figure=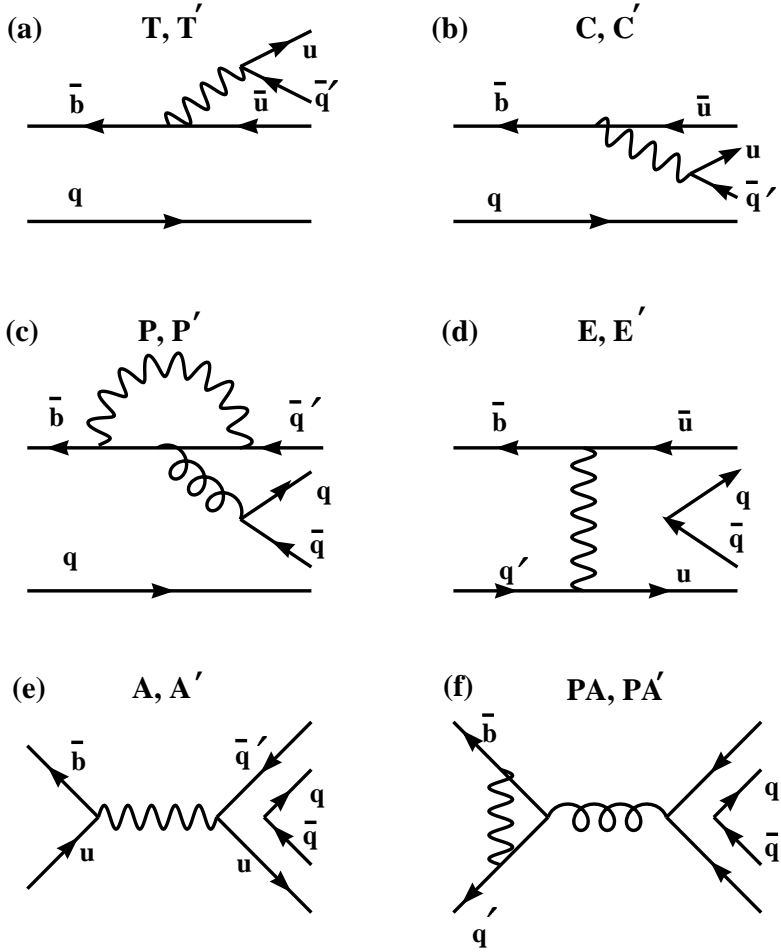,height=13.0cm,angle=0}}
\caption{Diagrams describing $B \to PP$ decays for $\Delta S = 0$ processes
(unprimed amplitudes) or $|\Delta S| = 1$ processes (primed amplitudes).
The $q$ quark denotes any member of the SU(3) triplet, $u,d,s$, whereas
$q'$ denotes $d$ or $s$.}
\label{figi-pp}
\end{figure}

\begin{figure}
\centerline{\psfig{figure=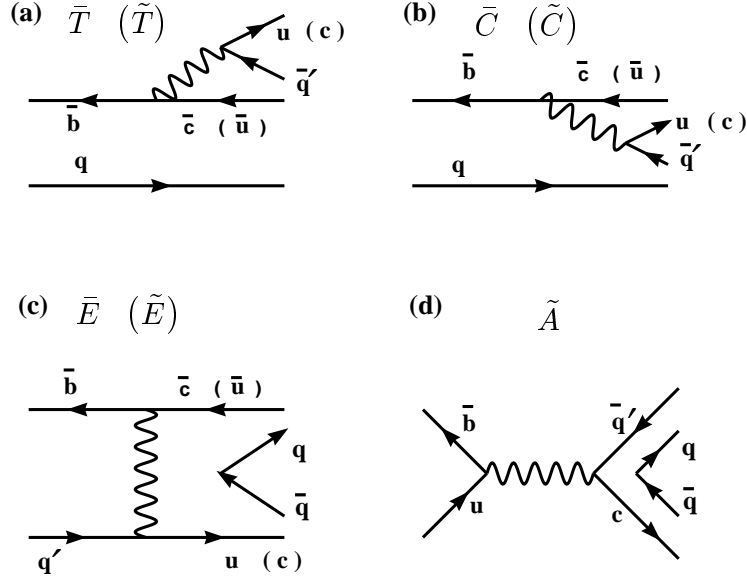,height=8.0cm,angle=0}}
\caption{Diagrams describing decays $B \to P \bar D$ or $B \to P D$ governed by
CKM factors $V_{cb}^* V_{ud}$ or $V_{cb}^*V_{us}$ ($\approx\lambda V_{cb}^*
V_{ud}$) (barred amplitudes), and $V_{ub}^* V_{cs}$ or $V_{ub}^* V_{cd}$
($\approx - \lambda V_{ub}^* V_{cs}$) (tilded amplitudes). The $q$ quark
denotes any member of the SU(3) triplet, $u,d,s$, whereas $q'$ denotes $d$ or
$s$.}
\label{figii-pd}
\end{figure}

\begin{figure}
\centerline{\psfig{figure=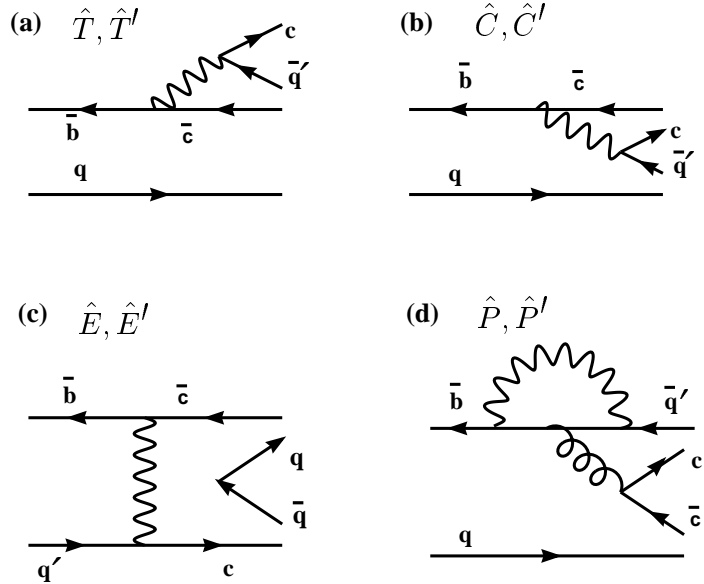,height=8.0cm,angle=0}}
\caption{Diagrams describing $B \to D \bar D$ (a,c,d), or $B \to \eta_c P$
(b) decays with $|\Delta S| = 1$ (unprimed amplitudes, $q'=s$) or with
$\Delta S = 0$ (primed amplitudes, $q'=d$). The $q$ quark denotes any
member $u,d,s$ of the SU(3) triplet.}
\label{figiii-dd}
\end{figure}

Diagrams describing $B$ decays are a particularly useful representation of
SU(3) amplitudes. There are six distinct diagrams, shown in
Fig.~\ref{figi-pp}. They consist of:
\begin{itemize}
\item a (color-favored) ``tree'' amplitude $T$,
\item a ``color-suppressed'' amplitude $C$,
\item a ``penguin'' amplitude $P$,
\item an ``exchange'' amplitude $E$,
\item an ``annihilation'' amplitude $A$, and
\item a ``penguin annihilation'' amplitude $PA$.
\end{itemize}
Of course, not all diagrams contribute to all classes of decays. In
particular,
\begin{enumerate}

\item All six diagrams contribute to the decays $B \to PP$ (see
Fig.~\ref{figi-pp}), but only five distinct linear combinations appear in
the amplitudes.

\item Three diagrams ($\bt$, $\bc$, $\be$) contribute to the decays $B \to
P \bar D$ (see Fig.~\ref{figii-pd}).

\item Four diagrams ($\tt$, $\tc$, $\te$, $\ta$) contribute to the decays
$B \to P D$ (see Fig.~\ref{figii-pd}).

\item Three diagrams ($\that$, $\phat$, $\ehat$) contribute to the decays
$B\to D\bar D$, but they only appear in two combinations ($\that+\phat$,
$\ehat$). Only one diagram ($\chat$) contributes to the decays $B\to\eta_c
P$ (see Fig.~\ref{figiii-dd}).

\end{enumerate}
As expected, one obtains the same number of diagrams (or combinations of
diagrams) contributing to the various classes of $B$ decays as was found
previously using group theory.

In Tables~\ref{tabi-pp}-\ref{tabviii-dd}, in the ``SU(3) invariant''
column, we present the decomposition in terms of diagrams of all the $B$
decays in the four classes:
\begin{enumerate}
\item $B \to PP$ (Tables~\ref{tabi-pp} and~\ref{tabii-pp}).
\item $B \to P \bar D$ (Tables~\ref{tabiii-pdbar} and~\ref{tabiv-pdbar}),
\item $B \to P D$ (Tables~\ref{tabv-pd} and~\ref{tabvi-pd}),
\item $B\to D\bar D$
        and $B\to\eta_c P$ (Tables~\ref{tabvii-dd} and~\ref{tabviii-dd}).
\end{enumerate}
Note that, for $B\to PP$, we include only the contributions {}from the $T$,
$C$ and $P$ diagrams. As discussed in \cite{BPP}, and reiterated in the
introduction, we expect the $E$, $A$ and $PA$ diagrams to be suppressed by
$f_B/m_B \approx 5\%$. We will be testing the validity of this assumption
with the $B\to P\bar D$ system. The decomposition of $B\to PP$ decays in
terms of all six diagrams can be found in~\cite{BPP}.
\bigskip

\renewcommand{\arraystretch}{1.2}

\begin{table}
\caption{Decomposition of $B \to PP$ amplitudes for $\Delta C = \Delta S =
0$ transitions in terms of graphical contributions shown in Fig.~1.
Amplitudes $E$, $A$, and $PA$ (and the corresponding SU(3)-breaking terms)
are neglected.}

\begin{center}
\begin{tabular}{l l c c c c c c} \hline
           &  Final        & SU(3)        & SU(3)   \\
           &  state        & invariant    & breaking  \\ \hline
$B^+\to$ & $\pi^+\pi^0$ & $-(T+C)/\s$ &  \\
         & $K^+ \bar K^0$ & $P$ & $P_3$   \\ \hline
$B^0\to$ & $\pi^+\pi^-$ & $-(T+P)$ &   \\
         & $\pi^0\pi^0$ & $-(C-P)/\s$ &  \\
         & $K^0 \bar K^0$ & $P$ & $P_3$ \\ \hline
$B_s\to$ & $\pi^+K^-$ & $-(T+P)$ & $-(T_2+P_2)$ \\
         & $\pi^0 \bar K^0$ & $-(C-P)/\s$ & $-(C_2-P_2)/\s$ \\ \hline
\end{tabular}
\end{center}
\label{tabi-pp}
\end{table}
\bigskip

\begin{table}
\caption{Decomposition of $B \to PP$ amplitudes for $\Delta C = 0,~ |\Delta
S| = 1$ transitions in terms of graphical contributions shown in
Fig.~1. Amplitudes $E'$, $A'$, and $PA'$ (and the corresponding
SU(3)-breaking terms) are neglected.}
\begin{center}
\begin{tabular}{l l c c c c c c} \hline
           &  Final        & SU(3)        & SU(3)   \\
           &  state        & invariant    & breaking  \\ \hline
$B^+\to$ & $\pi^+K^0$ & $P'$ & $P'_1$ \\
         & $\pi^0K^+$ & $-(T'+C'+P')/\s$ & $-(T'_1+C'_1+P'_1)/\s$ \\ \hline
$B^0\to$ & $\pi^-K^+$ & $-(T'+P')$ & $-(T'_1+P'_1)$ \\
         & $\pi^0K^0$ & $-(C'-P')/\s$ & $-(C'_1-P'_1)/\s$ \\  \hline
$B_s\to$ & $K^+K^-$ & $-(T'+P')$ & $-(T'_1+T'_2+P'_1+P'_2)$ \\
         & $K^0 \bar K^0$ & $P'$ & $P'_1+P'_2$ \\  \hline
\end{tabular}
\end{center}
\label{tabii-pp}
\end{table}
\bigskip

\begin{table}
\caption{Decomposition of amplitudes for processes governed by $V_{cb}^*
V_{ud} \sim {\cal O}(\lambda^2)$ in terms of graphical contributions shown
in Fig.~2.}
\begin{center}
\begin{tabular}{l c r r r} \hline
           &  Final              & SU(3)        & SU(3)   \\
           &  state              & invariant    & breaking  \\ \hline
$B^+ \to $ & $\pi^+ \bar D^0$    & $\bt+\bc$    & \\ \hline
$B^0 \to $ & $\pi^+ D^-$         & $\bt+\be$    & \\
           & $\pi^0 \bar D^0$    & $(\bc-\be)/\s$ & \\
           & $K^+ D_s^-$         &  $\be$       & $\be_{2}$ \\ \hline
$B_s \to $ & $\bar K^0 \bar D^0$ &  $\bc$       & $\bc_2$ \\
           & $\pi^+ D_s^-$       &  $\bt$       & $\bt_2$ \\ \hline
\end{tabular}
\end{center}
\label{tabiii-pdbar}
\end{table}

\begin{table}
\caption{Decomposition of amplitudes for processes governed by $V_{cb}^*
V_{us} \sim {\cal O}(\lambda^3)$ in terms of graphical contributions shown
in Fig.~2.}
\begin{center}
\begin{tabular}{l c r r r} \hline
           &  Final           & SU(3)           & SU(3)   \\
           &  state           & invariant       & breaking  \\ \hline
$B^+ \to $ & $K^+ \bar D^0$   & $\lambda(\bt+\bc)$ & $\lambda(\bt_1+\bc_1)$
\\ \hline
$B^0 \to $ & $K^+ D^-$        & $\lambda \bt$   & $\lambda \bt_1$  \\
           & $K^0 \bar D^0$   & $\lambda \bc$   & $\lambda \bc_1$  \\ \hline
$B_s \to $ & $\pi^+ D^-$      & $\lambda \be$   & $\lambda \be_{1}$   \\
           & $\pi^0 \bar D^0$ & $-\lambda \be/\s$ & $-\lambda \be_{1}/\s$ \\
           & $K^+ D_s^-$      & $\lambda(\bt+\be)$  & $\lambda
   (\bt_1+\bt_2+\be_1+\be_2)$ \\  \hline
\end{tabular}
\end{center}
\label{tabiv-pdbar}
\end{table}

\begin{table}
\caption{Decomposition of amplitudes for processes governed by $V_{ub}^*
V_{cs} \sim {\cal O}(\lambda^3)$ in terms of graphical contributions shown
in Fig.~2.}
\begin{center}
\begin{tabular}{l c r r r r} \hline
           &  Final        & SU(3)        & SU(3)   \\
           &  state        & invariant    & breaking  \\ \hline
$B^+ \to $ & $K^+ D^0$     &  $-\tc-\ta$  & $-\tc_1-\ta_1$  \\
           & $K^0 D^+$     &    $\ta$     & $\ta_1$   \\
           & $\pi^0 D_s^+$ &  $-\tt/\s$   & $-\tt_1/\s$  \\ \hline
$B^0 \to $ & $K^0 D^0$     &  $ -\tc$     & $-\tc_1$   \\
           & $\pi^- D_s^+$ &  $ -\tt$     & $-\tt_1$   \\ \hline
$B_s \to $ & $K^- D_s^+$   &  $ -\tt-\te$ & $-\tt_1-\tt_2-\te_1-\te_2$   \\
           & $\pi^- D^+$   &  $-\te$      & $-\te_{1}$ \\
           & $\pi^0 D^0$   &  $\te/\s$    & $\te_{1}/\s$   \\ \hline
\end{tabular}
\end{center}
\label{tabv-pd}
\end{table}

\begin{table}
\caption{Decomposition of amplitudes for processes governed by $V_{ub}^*
V_{cd} \sim {\cal O}(\lambda^4)$ in terms of graphical contributions shown
in Fig.~2.}
\begin{center}
\begin{tabular}{l c r r r r} \hline
           &  Final        & SU(3)        & SU(3)   \\
           &  state        & invariant    & breaking  \\ \hline
$B^+ \to $ & $\pi^+ D^0$   & $\lambda(\tc+\ta)$    & \\
           & $\pi^0 D^+$   & $\lambda(\tt-\ta)/\s$ & \\
        & $\bar K^0 D_s^+$ & $-\lambda\ta$        & $-\lambda\ta_2$ \\ \hline
$B^0 \to $ & $\pi^- D^+$   &  $\lambda(\tt+\te)$  & \\
           & $\pi^0 D^0$   &  $\lambda(\tc-\te)/\s$ &  \\
           & $K^- D_s^+$   &  $\lambda\te$       & $\lambda\te_{2}$  \\ \hline
$B_s \to $ & $K^- D^+$     &  $\lambda\tt$       & $\lambda\tt_2$  \\
        & $\bar K^0 D^0$   &  $\lambda\tc$       & $\lambda\tc_2$  \\ \hline
\end{tabular}
\end{center}
\label{tabvi-pd}
\end{table}
\bigskip

\begin{table}
\caption{Decomposition of amplitudes for $|\Delta S| = 1$ processes
involving a $c \bar c$ pair in the final state [leading behavior $\sim
{\cal O} (\lambda^2)$] in terms of graphical contributions shown in
Fig.~3.}
\begin{center}
\begin{tabular}{l c r r} \hline
           &  Final           & SU(3)           &  SU(3) \\
           &  state           & invariant       &  breaking \\ \hline
$B^+ \to $ & $D_s^+ \bar D^0$ & $\that + \phat$ & $\that_1 + \phat_1$ \\
           & $\eta_c K^+$     & $\chat$         & $\chat_1$   \\ \hline
$B^0 \to $ & $D_s^+ D^-$      & $\that + \phat$ & $\that_1 + \phat_1$ \\
           & $\eta_c K^0$     & $\chat$         & $\chat_1$   \\ \hline
$B_s \to $ & $D_s^+ D_s^-$    & $\that + \phat + \ehat $ &
              $\that_1 + \phat_1 + \ehat_1$ \\
           &                  &                          &
              $ + \that_2 + \phat_2 + \ehat_2$ \\
           & $D^+ D^-$        & $\ehat$         & $\ehat_{1}$ \\
           & $D^0 \bar D^0 $   & $-\ehat$        & $-\ehat_{1}$ \\ \hline
\end{tabular}
\end{center}
\label{tabvii-dd}
\end{table}

\begin{table}
\caption{Decomposition of amplitudes for $\Delta S = 0$ processes involving
a $c \bar c$ pair in the final state [leading behavior $\sim {\cal
O}(\lambda^3)$] in terms of graphical contributions shown in Fig.~3.}
\begin{center}
\begin{tabular}{l c r r} \hline
           &  Final           & SU(3)           &  SU(3) \\
           &  state           & invariant       &  breaking \\ \hline
$B^+ \to $ & $D^+ \bar D^0$ & $\that' + \phat'$ &   \\
           & $\eta_c \pi^+$   & $\chat'$        &   \\ \hline
$B^0 \to $ & $D^+ D^-$      & $\that' + \phat' + \ehat'$ &  \\
           & $D^0 \bar D^0$ & $-\ehat'$         &  \\
           & $D_s^+ D_s^-$  & $\ehat'$          & $\ehat'_{2}$ \\
           & $\eta_c \pi^0$ & $\chat'/\s$       &  \\ \hline
$B_s \to $ & $D^+ D_s^-$    & $\that' + \phat'$ & $\that'_2 + \phat'_2$ \\
           & $\eta_c \bar K^0$ & $\chat'$       & $\chat'_2$ \\ \hline
\end{tabular}
\end{center}
\label{tabviii-dd}
\end{table}
\newpage

\centerline{\bf III. SU(3)-BREAKING EFFECTS}
\bigskip

In the previous section, we discussed the decomposition of the various $B$
decays in terms of SU(3)-invariant amplitudes. We now turn to a discussion
of SU(3)-breaking effects.
\bigskip

\leftline{\bf A. SU(3)-breaking diagrams}
\bigskip

Flavor SU(3) is broken by the difference in the $u$, $d$ and $s$ quark
masses. Since the mass matrix transforms as a ${\bf 3} \times {\bf 3^*} =
{\bf 1} + {\bf 8}$ of SU(3), we use the octet piece to break SU(3) (the
singlet is, by definition, SU(3)-invariant). This breaking is first order
(i.e.\ linear) in the quark masses. In operator language, this corresponds
to the introduction of an operator $M$ into the SU(3)-invariant amplitudes,
in which $M$ is a linear combination of $\lambda_3$ and $\lambda_8$ (the
$\lambda_i$ are the usual Gell-Mann matrices). The $\lambda_3$ piece can be
neglected, since it corresponds to isospin breaking, which is expected to
be very small. We therefore have $M \sim \lambda_8$. It is now possible to
construct all SU(3)-breaking operators \`a la Savage and Wise \cite{SW},
and to examine their effects on $B$ decays.

\begin{figure}
\centerline{\psfig{figure=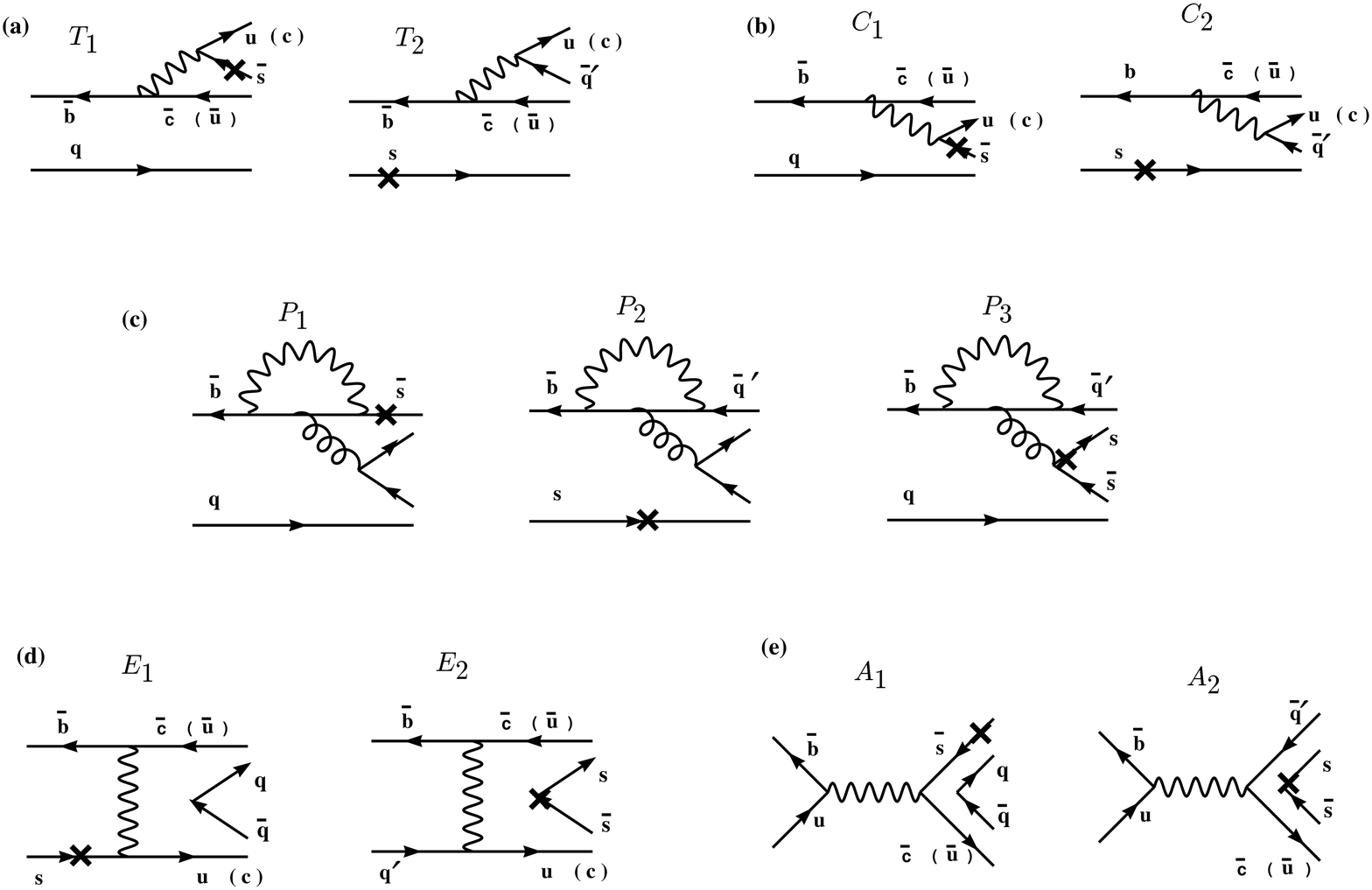,width=16cm}}
\caption{Diagrammatic representation of SU(3) symmetry breaking effects.
Crosses appear only on $s$ quark lines as explained in text.}
\label{figiv-xx}
\end{figure}

It is simpler, however, to think of the above in terms of a diagrammatic
decomposition of SU(3) amplitudes. It is the $s$-quark mass (or, more
precisely, the difference of the $s$-quark and the $d$-quark masses) which
breaks SU(3). The Gell-Mann matrix $\lambda_8 \sim {\rm diag}[1,1,-2]$ can be
written as the identity (which is SU(3)-invariant) plus the matrix ${\rm
diag}[0,0,-3]$. Thus, SU(3)-breaking operators will be nonzero only when an
$s$-quark is involved and an SU(3)-breaking diagram can be obtained {}from the
SU(3)-preserving diagrams of Fig.~\ref{figi-pp} by putting an ``X'' on any
$s$-quark line. The ``X'' corresponds to a mass-difference insertion
$(m_s-m_d)/\Lambda$, where $\Lambda$ is the scale of SU(3)
breaking. The SU(3)-breaking diagrams are shown in Fig.~\ref{figiv-xx}:
\begin{itemize}

\item There are two SU(3)-breaking diagrams which can be obtained {}from a
$T$ diagram: (1) in the $T_1$ diagram, the $s$-quark is among the decay
products of the $W$; (2) in the $T_2$ diagram, the $s$-quark is the
spectator quark.

\item There are two SU(3)-breaking diagrams which can be obtained {}from a
$C$ diagram: (1) in the $C_1$ diagram, the $s$-quark is among the decay
products of the $W$; (2) in the $C_2$ diagram, the $s$-quark is the
spectator quark.

\item There are three SU(3)-breaking diagrams which can be obtained {}from
a $P$ diagram: (1) in the $P_1$ diagram, there is a $b\to s$ transition;
(2) in the $P_2$ diagram, the $s$-quark is the spectator quark; (3) in the
$P_3$ diagram, an $s\bar s$ quark pair is created.

\item There are two SU(3)-breaking diagrams which can be obtained {}from a
$E$ diagram: (1) in the $E_{1}$ diagram, the $s$-quark is in the decaying
($B_s$) meson; (2) in the $E_{2}$ diagram, an $s\bar s$ quark pair is
created.

\item There are two SU(3)-breaking diagrams which can be obtained {}from a
$A$ diagram: (1) in the $A_1$ diagram, the $s$-quark is among the decay
products of the $W$; (2) in the $A_2$ diagram, an $s\bar s$ quark pair is
created.

\item There are two SU(3)-breaking diagrams which can be obtained {}from a
$PA$ diagram. They are not shown in Fig.~\ref{figiv-xx} since we will never
make use of them. However we list them here for completeness: (1) in the
$PA_1$ diagram, the $s$-quark is in the decaying ($B_s$) meson; (2) in the
$PA_2$ diagram, an $s\bar s$ quark pair is created.

\end{itemize}

It is now straightforward to establish which SU(3)-breaking diagrams
contribute to the various $B$ decays:

\begin{enumerate}

\item $B \to PP$: all six diagrams contribute to these decays (albeit in
only five distinct linear combinations), so all 13 SU(3)-breaking diagrams
will contribute, though only in 10 distinct linear combinations.

\item $B \to P \bar D$: The $\bt$, $\bc$ and $\be$ diagrams contribute to
these decays, so there are six possible SU(3)-breaking contributions
($\bt_1$, $\bt_2$, $\bc_1$, $\bc_2$, $\be_1$, $\be_2$).

\item $B \to P D$: The $\tt$, $\tc$, $\te$ and $\ta$ diagrams contribute to
these decays, so there are eight possible SU(3)-breaking contributions
($\tt_1$, $\tt_2$, $\tc_1$, $\tc_2$, $\te_1$, $\te_2$, $\ta_1$, $\ta_2$).

\item $B\to D\bar D$: Two combinations ($\that+\phat$, $\ehat$) of the
three diagrams $\that$, $\phat$ and $\ehat$ contribute to these decays, so
there are four possible SU(3)-breaking contributions ($\that_1 + \phat_1$,
$\that_2 + \phat_2$, $\ehat_1$, $\ehat_2$). ($\phat_3$ never appears since
in these decays a $c\bar c$ quark pair is created, not an $s\bar s$ pair.)
\\
$B\to\eta_c P$: The $\chat$ diagram contributes to these decays, so there
are two possible SU(3)-breaking contributions ($\chat_1$, $\chat_2$).

\end{enumerate}

In Tables~\ref{tabi-pp}-\ref{tabviii-dd}, in the ``SU(3) breaking'' column,
we present the SU(3)-breaking contributions to all the $B$ decays in the
four classes:
\begin{enumerate}
\item $B \to PP$ (Tables~\ref{tabi-pp} and~\ref{tabii-pp}).
\item $B \to P \bar D$ (Tables~\ref{tabiii-pdbar} and~\ref{tabiv-pdbar}),
\item $B \to P D$ (Tables~\ref{tabv-pd} and~\ref{tabvi-pd}),
\item $B\to D\bar D$ and $B\to\eta_c P$
     (Tables~\ref{tabvii-dd} and~\ref{tabviii-dd}),
\end{enumerate}
For $B\to PP$, we include only the SU(3)-breaking contributions which are
derived {}from the $T$, $C$ and $P$ diagrams and their primed counterparts.
Those SU(3)-breaking diagrams which are related to the $E$, $A$ and $PA$
diagrams are expected to be much smaller (see below).

Note that, in $T$-type diagrams, the weak current is coupled directly to a
final-state meson (see Figs.~\ref{figi-pp}--\ref{figiii-dd}). Therefore,
assuming factorization, SU(3)-breaking effects in the decay of the $W$ can
be directly related to meson decay constants. Specifically, for $B\to PP$
and $B\to P\bar D$,
\beq
\left\vert {T + T_1\over T}\right\vert = {f_K\over f_\pi}~,
\label{fkfpi}
\eeq
while for $B\to PD$ and $B\to D\bar D$ we have
\beq
\left\vert {T + T_1\over T}\right\vert = {f_{D_s} \over f_D}~.
\label{fdsfd}
\eeq
(In the above, the symbol ``$T$'' represents any $T$-type diagram in
Figs.~\ref{figi-pp}--\ref{figiii-dd}.)

Since the $T_2$, $C_2$ and $P_2$ corrections involve the spectator quark,
these can be interpreted as form-factor corrections. The remaining
SU(3)-breaking corrections are related to the difference in the production
amplitudes for $s \bar s$ and $u \bar u$ ($d \bar d$).

In all cases, the SU(3)-breaking diagrams may have different strong phases
than the parent diagrams, so that final-state phases can be affected. In
particular, in Eqs.~(\ref{fkfpi}) and (\ref{fdsfd}) above, the quantity
$1+T_1/T$ is in general equal to the ratio of decay constants times an
unknown phase.

\bigskip
\leftline{\bf B. Expected sizes of the various diagrams}
\bigskip

Not all of the SU(3)-invariant contributions are expected to be equally
large -- we expect there to be a range of magnitudes. The SU(3)-violating
contributions should obey a similar hierarchy.

For example, the $T$, $C$, $E$ and $A$ contributions to a particular decay
all have the same CKM matrix elements. However, for dynamical reasons, the
$T$ diagram is expected to dominate. The $C$ diagram is color-suppressed,
so naively its magnitude should be smaller than that of the $T$ diagram by
a factor 1/3. Model calculations suggest that the ratio $|C/T|$ is in fact
somewhat smaller, about 0.2 \cite{BSW,Alam}. For the purposes of
comparison, we will take $|C/T| \sim \lambda$. (Note that the use of the
parameter $\lambda$ here is not related in any way to the CKM matrix
elements of $C$ and $T$ -- it is simply used to keep track of the relative
size of the two diagrams.) As previously mentioned, the $E$ and $A$
diagrams are expected to be suppressed relative to the $T$ diagrams by a
factor $f_B/m_B \approx 5\% \sim \lambda^2$. (Again, the parameter
$\lambda$ is used here only as an approximate measure of the relative
size.) Thus, the approximate relative sizes of these four SU(3)-invariant
contributions are $|T|:|C|:|E|,|A| = 1:\lambda:\lambda^2$.

We do not know how large the SU(3)-breaking effects are. Our one clue comes
{}from the ratio $f_K/f_\pi = 1.2$, which appears naturally if
factorization is assumed, i.e.\ $(f_K - f_\pi)/f_\pi \approx 0.2 \sim
\lambda$. Assuming all SU(3)-breaking effects are of this order, we expect
$|T_i/T| \sim \lambda$, $|C_i/C| \sim \lambda$, etc. (If the SU(3)-breaking
effects should turn out to be significantly larger, then our lowest-order
parametrization of SU(3) breaking would probably be suspect.)

The $P$ and $PA$ contributions have to be considered separately, since they
have different CKM matrix elements than the $T$, $C$, $E$ and $A$ diagrams.
We will discuss them as they arise in the various $B$ decays below.

{\it A word of caution to the reader}: In what follows, we estimate the
relative sizes (in powers of $\lambda$) of the SU(3)-invariant and
SU(3)-breaking diagrams which contribute to all two-body hadronic $B$
decays. In later sections we often use this estimated hierarchy to isolate
the largest effect in a particular decay (including appropriate
explanations, of course). However, one must be careful not to take this
hierarchy too literally. Not only are these only educated guesses, but
$\lambda$ is not that small a number -- a factor of 4 enhancement or
suppression can easily turn an effect of ${\cal O}(\lambda^n)$ into an
effect of ${\cal O}(\lambda^{n\pm 1})$.

\medskip

\noindent
{\it 1. $B \to PP$ decays:} For these decays, the $\bar b \to \bar u u \bar
d$ and $\bar b \to \bar u u \bar s$ transitions must be analysed
separately, since the penguin contributions play a different role in the
two cases.

The dominant diagram in $\bar b \to \bar u u \bar d$ decays is $T$, whose
CKM matrix elements are $V_{ub}^* V_{ud}$. Based on the above discussion,
relative to $|T|$ we expect that $|C|$, $|E|$, $|A|$, and the SU(3)
corrections to $T$, $C$, $E$ and $A$ are suppressed by various powers of
$\lambda$. The $P$ diagram is also smaller than the $T$ diagram, but its
suppression factor is more uncertain. The CKM matrix elements for $P$ are
$V_{tb}^* V_{td}$ \cite{burasfleischer}. Although $|V_{td}| > |V_{ub}|$,
there are suppressions due to the loop and to $\alpha_s(m_b)\simeq 0.2$.
Allowing for the possibility that the $P$ matrix elements are enhanced
relative to the $T$ matrix elements, a conservative estimate is $|P/T| \sim
{\cal O}(\lambda)$ (although this is likely to be somewhat smaller
\cite{penguinsize}). The $PA$ diagram should be suppressed relative to the
$P$ diagram by a factor $f_B/m_B \sim \lambda^2$. Thus, for $\Delta C =
\Delta S = 0$ transitions, relative to the dominant $|T|$ contribution we
expect the following approximate hierarchy to hold:
\newpage
\begin{eqnarray}
\label{buudhierarchy}
{\cal O}(\lambda^0) & : & |T|
\nonumber \\
{\cal O}(\lambda) & : & |C|,~|P|,~{\hbox{SU(3) corrections to $T$,}}
\nonumber \\
{\cal O}(\lambda^2) & : & |E|,~|A|,~{\hbox{SU(3) corrections to $C$ and
$P$,}} \nonumber \\
{\cal O}(\lambda^3) & : & |PA|,~{\hbox{SU(3) corrections to $E$ and $A$,}}
\\
{\cal O}(\lambda^4) & : & {\hbox{SU(3) corrections to $PA$.}} \nonumber
\end{eqnarray}
This implies that, if one neglects the $E$ and $A$ contributions to such
decays, it is consistent to also ignore all SU(3)-breaking effects except
the corrections to $T$.

For $\bar b \to \bar u u \bar s$ transitions, the relevant CKM matrix
elements in a $T'$ diagram are $V_{ub}^* V_{us} \sim {\cal O}(\lambda^4)$,
while those for the $P'$ diagram (which corresponds to a $\bar b \to \bar
s$ transition) are $V_{tb}^* V_{ts} \sim {\cal O}(\lambda^2)$. There is a
suppression for the $P'$ diagram due to the loop and to $\alpha_s(m_b)$,
and we estimate this as above to be ${\cal O}(\lambda)$. The conclusion is
that, in these decays, it is the $P'$ diagram which dominates. Thus, for
$\Delta C = \Delta S = 0$ transitions, relative to $|P'|$ we expect the
following approximate hierarchy of contributions:
\begin{eqnarray}
\label{buushierarchy}
{\cal O}(\lambda^0) & : & |P'|
\nonumber \\
{\cal O}(\lambda) & : & |T'|,~{\hbox{SU(3) corrections to $P'$,}}
\nonumber \\
{\cal O}(\lambda^2) & : & |C'|,~|PA'|,~{\hbox{SU(3) corrections to $T'$}}
\nonumber \\
{\cal O}(\lambda^3) & : & |E'|,~|A'|,~{\hbox{SU(3) corrections to $C'$ and
$PA'$,}}  \\
{\cal O}(\lambda^4) & : & {\hbox{SU(3) corrections to $E'$ and $A'$.}}
\nonumber
\end{eqnarray}
It should be stressed, however, that this estimated hierarchy is on less
solid ground than that for $\bar b \to \bar u u \bar d$ transitions, since
our knowledge of penguin contributions to hadronic $B$ decays is rather
sketchy at the moment. However, if this hierarchy holds, then it is
probably consistent to ignore the $C'$ contribution in Table
\ref{tabii-pp}, as well as all SU(3)-breaking effects except the $P_i'$.
\smallskip

\noindent
{\it 2. $B \to P \bar D$ decays:} The largest contribution to these decays
is $\bt$. Relative to $|\bt|$ we expect that $|\bc|$, $|\bt_1|$ and
$|\bt_2|$ are ${\cal O}(\lambda)$; $|\be|$, $|\bc_1|$ and $|\bc_2|$ are
${\cal O}(\lambda^2)$; and $|\be_1|$ and $|\be_2|$ are ${\cal
O}(\lambda^3)$.
\smallskip

\noindent
{\it 3. $B \to P D$ decays:} The largest contribution to these decays is
$\tt$. Relative to $|\tt|$ we expect that $|\tc|$, $|\tt_1|$ and $|\tt_2|$
are ${\cal O}(\lambda)$; $|\te|$, $|\ta|$, $|\tc_1|$ and $|\tc_2|$ are
${\cal O}(\lambda^2)$; and $|\te_1|$, $|\te_2|$, $|\ta_1|$ and $|\ta_2|$
are ${\cal O}(\lambda^3)$.
\smallskip

\noindent
{\it 4. $B \to D \bar D$ and $B \to \eta_c P$ decays:}  For the $B \to D
\bar D$ decays, the $\that$ diagram dominates, and the $\ehat$ diagram is
suppressed relative to it by a factor of ${\cal O}(\lambda^2)$. As for the
$\phat$ diagram, its CKM matrix elements are about the same size as those
of $\that$, but there are suppressions due to the loop, to
$\alpha_s(m_b)\simeq 0.2$, and to the fact that a $c\bar c$ pair must be
created. Taking all factors into account, the total suppression is probably
of ${\cal O}(\lambda^2)$, stronger than that in $B\to PP$ decays. With this
assumption, relative to $|\that|$ we expect that $|\that_1|$ and
$|\that_2|$ are ${\cal O}(\lambda)$, $|\ehat|$ and $|\phat|$ are ${\cal
O}(\lambda^2)$, and $|\ehat_1|$, $|\ehat_2|$, $|\phat_1|$ and $|\phat_2|$
are ${\cal O}(\lambda^3)$. For $B \to \eta_c P$ decays, the $\chat$ diagram
dominates, and the $|\chat_1|$ and $|\chat_2|$ corrections are suppressed
relative to it by a factor of ${\cal O}(\lambda)$.

\bigskip
\centerline{\bf IV. TESTS FOR SU(3) BREAKING AND NEGLECT OF $E$, $A$
DIAGRAMS}
\bigskip

We now inspect Tables~\ref{tabi-pp}-\ref{tabviii-dd} for relations which
test for the magnitude of SU(3)-breaking terms and for the absence of $E$
and $A$ diagrams. We consider pairs of tables together, since they are
generally related by a factor $\lambda$. We first discuss relations which
are expected to hold in the presence of SU(3) breaking, usually as a
consequence of the isospin properties of the weak Hamiltonian. We then
discuss general tests for SU(3) breaking, keeping $E$ and $A$
contributions, and finally note the additional relations which follow if
such terms are neglected. In what follows we shall always work to first
order in SU(3) breaking. We remind the reader that, aside {}from the
decays $B \to PP$, one is free to change one or both final-state
pseudoscalar mesons to a vector meson in all the relations to be quoted
below.

\bigskip
\leftline{\bf A. $B \to PP$ decays}
\bigskip

We refer the reader to \cite{BPP} for our conventions regarding identical
particles. Amplitudes are defined in such a way that their squares always
yield decay rates with the same constant of proportionality.
\smallskip

\noindent
{\it 1. Relations following merely {}from isospin} consist of the equality
\beq
A(B_s \to \pi^+ \pi^-) = -\s A(B_s \to \pi^0 \pi^0)~~~,
\eeq
the triangle relation
\beq \label{eqn:ispintri}
\s A(B^+ \to \pi^+ \pi^0) = A(B^0 \to \pi^+ \pi^-) + \s A(B^0 \to
\pi^0 \pi^0)~~~,
\eeq
and the quadrangle relation \cite{pik}
\beq
\label{quadrangle}
A(B^+ \to \pi^+ K^0) + \s A(B^+ \to \pi^0 K^+)
= A(B^0 \to \pi^- K^+) + \s A(B^0 \to \pi^0 K^0)~~~.
\eeq

\noindent
{\it 2. Within SU(3) symmetry, $B^+$ decays to $\pi \pi$ and $\pi K$ are
related} \cite{BPP,PRL,PLB}:
\beq \label{eqn:su3tri}
A(B^+ \to \pi^+ K^0) + \s A(B^+ \to \pi^0 K^+) = \lambda \s A(B^+ \to \pi^+
\pi^0)~~~.
\eeq

The general treatment of SU(3) breaking for $B \to PP$ decays (including
$E$, $A$, and $PA$ terms) involves a large number of contributions, since
all the quarks in the final state transform as flavor triplets or
antitriplets. In the remaining relations, based on Tables~\ref{tabi-pp}
and~\ref{tabii-pp}, we ignore the effects of $E$, $A$, and $PA$ and the
corresponding SU(3)-breaking terms. Numerous tests for the presence of $E$,
$A$, and $PA$ were suggested in \cite{BPP}. Even with this simplification,
we find that SU(3)-breaking effects are harder to separate {}from one
another than in the cases involving one or more charmed quarks in the final
state. We find the following relations:
\smallskip

\noindent
{\it 3. One amplitude relation} is preserved in the presence of SU(3)
breaking:
\beq
A(B^+ \to K^+ \bar K^0) = A(B^0 \to K^0 \bar K^0)~~~.
\eeq
Both amplitudes are $P + P_3$. This relation would not necessarily hold in
the presence of unequal $A$ and $PA$ contributions, since the left-hand
side receives a contribution $A$ while the right-hand side has an
additional $PA$ term \cite{BPP}.

\smallskip
\noindent
{\it 4. Several combinations of SU(3)-breaking terms} can be extracted
{}from the data:
\begin{eqnarray}
\label{su3one}
\Gamma(B_s \to \pi^+ K^-)/\Gamma(B^0 \to \pi^+ \pi^-)
& = & 1 + 2 {\rm~Re}[(T_2 + P_2)/(T + P)]~~~, \\
\label{su3two}
\Gamma(B_s \to \pi^0 \bar K^0)/\Gamma(B^0 \to \pi^0 \pi^0)
& = & 1 + 2 {\rm~Re}[(C_2 - P_2)/(C - P)]~~~, \\
\label{su3three}
\Gamma(B_s \to K^0 \bar K^0)/\Gamma(B^+ \to \pi^+ K^0)
& = & 1 + 2 {\rm~Re}(P_2'/P')~~~, \\
\label{su3four}
\Gamma(B_s \to K^+ K^-)/\Gamma(B^0 \to \pi^- K^+)
& = & 1 + 2 {\rm~Re}[(T_2' + P_2')/(T' + P')]~~~.
\end{eqnarray}
(Only the real parts of the SU(3)-breaking terms appear here and below,
since we are working only to linear order in these terms.) Our program of
ignoring $E$, $A$ and $PA$ terms is equivalent to keeping only the
lowest-order corrections to the dominant term in any decay. If our
estimates (see Sec.\ III B) of the approximate sizes of the various
SU(3)-breaking terms are correct, $C_2$, $P_2$ and $T_2'$ are negligible to
the order at which we are working. Furthermore, in the SU(3) corrections on
the right-hand sides of the above equations, we need only keep the largest
terms in both the numerator and denominator. The other contributions are
subdominant and can be ignored. Thus, at this level of approximation the
SU(3)-breaking quantity that is measured in Eq.~(\ref{su3one}) above is
${\rm Re}(T_2/T)$, while the quantity ${\rm Re}(P_2'/P')$ is measured in
both Eqs.~(\ref{su3three}) and (\ref{su3four}). To this order, since we
have neglected $E$ and $PA$ terms in the denominator of the left-hand side
of Eq.~(\ref{su3two}) which are of the same order as SU(3)-breaking terms,
the SU(3)-breaking factor on the right-hand side should be ignored; we
cannot say anything about ${\cal O}(\lambda)$ corrections in this case.

SU(3)-breaking terms modify the triangle relation (\ref{eqn:su3tri}):
\beq
A(B^+ \to \pi^+ K^0) + \s A(B^+ \to \pi^0 K^+) =
\left( 1 + \frac{T_1' + C_1'} {T' + C'} \right )
\lambda \s A(B^+ \to \pi^+ \pi^0)~~~.
\eeq
We have argued above that the $C_1'$ and $C'$ terms in the above expression
give subdominant SU(3) corrections, and are therefore negligible. Thus,
using Eq.~(\ref{fkfpi}), we see that the SU(3)-breaking effect which enters
the relation between the $I = 3/2$ $B \to \pi K$ amplitude and $\lambda$
times the $I = 2$ $B \to \pi \pi$ amplitudes is just $f_K/f_\pi$ (times a
possible strong phase). This is, in fact, what we estimated previously
\cite{BPP}.

To relate various contributions in $B \to \pi \pi$ and $B \to \pi K$ decays
to one another, Silva and Wolfenstein \cite{SilWo} neglected $E$ and $PA$
in $B^0 \to \pi^+ \pi^-$ and assumed that $T_1'/T' = P_1'/P'$ in $B^0 \to
\pi^- K^+$. We find that this assumption is difficult to test using the
decays of Tables~\ref{tabi-pp} and~\ref{tabii-pp}.

\bigskip
\leftline{\bf B. $B \to P \bar D$ decays}
\bigskip

\noindent
{\it 1. An isospin amplitude relation} connects the amplitudes for $B_s
\to \pi \bar D$:
\beq
A(B_s \to \pi^+ D^-) = - \s A(B_s \to \pi^0 \bar D^0)~~~.
\eeq

\noindent
{\it 2. Isospin triangle relations} connect the amplitudes for
$B \to \pi \bar D$:
\beq \label{eqn:pid}
A(B^+ \to \pi^+ \bar D^0) = A(B^0 \to \pi^+ D^-) + \s A(B^0 \to \pi^0 \bar
D^0) ~~.
\eeq
and the amplitudes for $B \to K \bar D$:
\beq \label{eqn:kd}
A(B^+ \to K^+ \bar D^0) = A(B^0 \to K^+ D^-) + A(B^0 \to K^0 \bar D^0)~~~.
\eeq

\noindent
{\it 3. One relation among six amplitudes} holds in the presence of
first-order SU(3) breaking when $E$ terms are retained:
$$
A(B_s \to K^+ D_s^-) - A(B_s \to \pi^+ D^-) - A(B^0 \to K^+ D^-)
$$
\beq \label{eqn:hex1}
= \lambda [A(B_s \to \pi^+ D_s^-) + A(B^0 \to K^+ D_s^-) - A(B^0 \to \pi^+
D^-)]~~~.
\eeq

We can also use the results of these two tables to learn about the sizes of
the SU(3) breaking:
\smallskip

\noindent
{\it 4. The real part of the ratio $(\bt_1 + \bc_1)/(\bt + \bc)$} may be
learned {}from the ratio
\beq
\Gamma(B^+ \to K^+ \bar D^0)/ \lambda^2 \Gamma(B^+ \to \pi^+ \bar D^0)
= 1 + 2 {\rm~Re} [(\bt_1 + \bc_1)/(\bt + \bc)]~~~.
\eeq
One must write this relation in terms of the real part of the ratio of the
SU(3) breaking and SU(3) invariant terms since strong final-state phases
may not be the same in the $K^+ \bar D^0$ and $\pi^+ \bar D^0$ channels.
Once again, if our estimates of the approximate sizes of the SU(3)-breaking
terms are correct, the $\bc_1$ and $\bc$ terms in the above expression are
negligible since they are simply higher-order corrections. In this case the
above rate ratio is simply equal to $(f_K/f_\pi)^2$ [see
Eq.~(\ref{fkfpi})].
\smallskip

\noindent {\it 5. Other rate ratios} provide information on combinations of
parameters:
\beq
\Gamma(B^0 \to K^0 \bar D^0)/ \lambda^2 \Gamma(B_s \to \bar K^0 \bar D^0)
= 1 + 2 {\rm~Re} [(\bc_1 - \bc_2)/\bc]~~~,
\eeq
\beq
\Gamma(B_s \to \pi^+ D^-)/ \lambda^2 \Gamma(B^0 \to K^+ D_s^-)
= 1 + 2 {\rm~Re} [(\be_1 - \be_2)/\be]~~~.
\eeq

If we now neglect all $\be$ contributions (there are no $\ba$ terms in $B
\to P \bar D$ decays),
\smallskip

\noindent
{\it 6. Three decay rates vanish:}
\beq
\Gamma(B^0 \to K^+ D_s^-) = \Gamma(B_s \to \pi^+ D^-) = \Gamma(B_s \to
\pi^0 \bar D^0) = 0~~~.
\eeq
Upper limits on the size of $\be$ terms can already be obtained from the
data: $B(B^0\to K^+ D_s^{*-})/B(B^0 \to \pi^+ D^{*-}) < 1/12$ and $B(B^0\to
K^+ D_s^-)/B(B^+ \to \pi^+ \bar D^0) < 1/20$ \cite{Alex}. Of course, the
measurement of these ratios will have to improve by more than an order of
magnitude in order to detect $\be$ effects at the expected level, but it is
interesting that we already have significant experimental evidence
regarding the suppression of the $\be$ terms. We will discuss the
experimental data further in Sec.~VII.

In addition we learn more about the SU(3)-breaking terms:
\smallskip

\noindent
{\it 7. The real parts of $\bt_{1,2}/\bt$ and $\bc_{1,2}/\bc$} can be
learned separately {}from the ratios
\beq
\Gamma(B^0 \to K^+ D^-)/\lambda^2 \Gamma(B^0 \to \pi^+ D^-)
= 1 + 2 {\rm~Re}(\bt_1/\bt)~~~,
\eeq
\beq \label{eqn:ttone}
\Gamma(B_s \to \pi^+ D_s^-)/\Gamma(B^0 \to \pi^+ D^-)
= 1 + 2 {\rm~Re}(\bt_2/\bt)~~~,
\eeq
\beq \label{eqn:tttwo}
\Gamma(B_s \to K^+ D_s^-)/\Gamma(B^0 \to K^+ D^-)
= 1 + 2 {\rm~Re}(\bt_2/\bt)~~~,
\eeq
\beq
\Gamma(B^0 \to K^0 \bar D^0)/ 2 \lambda^2 \Gamma(B^0 \to \pi^0 \bar D^0)
= 1 + 2 {\rm~Re}(\bc_1/\bc)~~~,
\eeq
\beq
\Gamma(B_s \to \bar K^0 \bar D^0)/2 \Gamma(B^0 \to \pi^0 \bar D^0)
= 1 + 2 {\rm~Re}(\bc_2/\bc)~~~.
\eeq
Using Eq.~\ref{fkfpi}, the first relation above is in fact equal to
$(f_K/f_\pi)^2$. Furthermore, if our estimated hierarchy is correct, the
$\bc_i$ terms are about the same size as the $\be$ terms which we have
neglected. Therefore, to this order, the last two relations are reliable only
up to ${\cal O}(1)$, not to ${\cal O}(\lambda)$.

\noindent
{\it 8. A consistency check} may be performed by comparing the results of
Eqs.~(\ref{eqn:ttone}) and (\ref{eqn:tttwo}).
\bigskip

\leftline{\bf C. $B \to P D$ decays}
\bigskip

Since these processes are at most of order $\lambda^3$, they will be less
valuable for testing SU(3) breaking and neglect of $E$ than the $B \to P
\bar D$ decays mentioned above. These processes do provide a unique testing
ground for the presence of $A$ contributions, however. Moreover, the ratio
$\tc/\tt$ of color-suppressed to color-non-suppressed amplitudes (which may
differ {}from the corresponding ratio $\bc/\bt$ for $B \to P \bar D$
decays) is important for the measurement \cite{MG} of the weak phase
$\gamma$ using $B \to K D_{\rm CP}$ decays.

\smallskip
\noindent
{\it 1. Two isospin relations} between amplitudes exist:
\beq
\s A(B^+ \to \pi^0 D_s^+) = A(B^0 \to \pi^- D_s^+)~~~,
\eeq
\beq
A(B_s \to \pi^- D^+) = - \s A(B_s \to \pi^0 D^0)~~~.
\eeq

\noindent
{\it 2. One isospin triangle} can be found:
\beq \label{eqn:bpdtri}
A(B^+ \to K^+ D^0) + A(B^+ \to K^0 D^+) = A(B^0 \to K^0 D^0)~~~.
\eeq

\noindent
{\it 3. One isospin quadrangle} holds:
\beq \label{eqn:bpdquad}
A(B^+ \to \pi^+ D^0) + \s A(B^+ \to \pi^0 D^+)
= A(B^0 \to \pi^- D^+) + \s A(B^0 \to \pi^0 D^0)~~~.
\eeq

\noindent
{\it 4. One relation among six amplitudes} is valid in the presence of all
first-order terms:
$$
A(B^0 \to \pi^- D^+) - A(B^0 \to K^- D_s^+) - A(B_s \to K^- D^+)
$$
\beq \label{eqn:hex2}
= \lambda [ A(B_s \to K^- D_s^+) - A(B_s \to \pi^- D^+) - A(B^0 \to \pi^-
D_s^+) ]~~~.
\eeq

We also obtain a number of additional results:
\smallskip

\noindent
{\it 5. The following SU(3)-breaking terms} can be extracted {}from ratios
of rates:
\beq
\lambda^2 \Gamma(B^+ \to K^+ D^0)/ \Gamma(B^+ \to \pi^+ D^0)
= 1 + 2 {\rm~Re}[(\tc_1 + \ta_1)/(\tc + \ta)]~~~,
\eeq
\beq
\lambda^2 \Gamma(B^+ \to K^0 D^+)/ \Gamma(B^+ \to \bar K^0 D_s^+)
= 1 + 2 {\rm~Re}[(\ta_1 - \ta_2)/\ta]~~~,
\eeq
\beq
\lambda^2 \Gamma(B^0 \to K^0 D^0)/ \Gamma(B_s \to \bar K^0 D^0)
= 1 + 2 {\rm~Re}[(\tc_1 - \tc_2)/\tc]~~~,
\eeq
\beq
\lambda^2 \Gamma(B^0 \to \pi^- D_s^+) / \Gamma(B_s \to K^- D^+)
= 1 + 2 {\rm~Re}[(\tt_1 - \tt_2)/\tt]~~~,
\eeq
\beq
\lambda^2 \Gamma(B_s \to \pi^- D^+) / \Gamma(B^0 \to K^- D_s^+)
= 1 + 2 {\rm~Re}[(\te_1 - \te_2)/\te]~~~.
\eeq

Now we examine the consequence of neglecting $\te$ and $\ta$ contributions.
\smallskip

\noindent
{\it 6. The following 5 rates vanish:}
$$
\Gamma(B^+ \to K^0 D^+) = \Gamma(B_s \to \pi^- D^+) = \Gamma(B_s \to \pi^0
D^0)
$$
\beq
= \Gamma(B^+ \to \bar K^0 D_s^+) = \Gamma(B^0 \to K^- D_s^+) = 0~~~.
\eeq

The vanishing of the rate for $B^+ \to K^0 D^+$ implies, through the
isospin triangle (\ref{eqn:bpdtri}), a relation between amplitudes with
isospins 0 and 1 in the direct channel, and the equality of the amplitudes
for $B^+ \to K^+ D^0$ and $B^0 \to K^0 D^0$. Since these processes are
color-suppressed, the violation of the rate relation $\Gamma(B^+ \to K^+
D^0) = \Gamma(B^0 \to K^0 D^0)$ would probably be the most stringent test
we could devise for the presence of annihilation ($\ta$) contributions.
\smallskip

\noindent
{\it 7. The one quadrangle relation (\ref{eqn:bpdquad}) becomes two
amplitude relations:}
\beq
A(B^+ \to \pi^+ D^0) = \s A(B^0 \to \pi^0 D^0)~~~,
\eeq
\beq
\s A(B^+ \to \pi^0 D^+) = A(B^0 \to \pi^- D^+)~~~.
\eeq

\noindent
{\it 8. In addition the following SU(3)-breaking terms can be extracted:}
\beq
\lambda^2 \Gamma(B^+ \to K^+ D^0) / \Gamma(B^+ \to \pi^+ D^0)
= 1 + 2 {\rm~Re}(\tc_1/\tc)~~~,
\eeq
\beq
\Gamma(B_s \to \bar K^0 D^0)/\Gamma(B^+ \to \pi^+ D^0)
= 1 + 2 {\rm~Re}(\tc_2/\tc)~~~,
\eeq
\beq
\lambda^2 \Gamma(B^0 \to \pi^- D_s^+) / \Gamma(B^0 \to \pi^- D^+)
= 1 + 2 {\rm~Re}(\tt_1/\tt)~~~,
\eeq
\beq
\Gamma(B_s \to K^- D_s^+)/\Gamma(B^0 \to \pi^- D_s^+)
= 1 + 2 {\rm~Re}(\tt_2/\tt)~~~.
\eeq
\beq
\Gamma(B_s \to K^- D^+)/\Gamma(B^0 \to \pi^- D^+)
= 1 + 2 {\rm~Re}(\tt_2/\tt)~~~.
\eeq
If the $\tc_i$ terms are of the same order as the $\te$ and $\ta$ terms, as
we expect, the first two of the above rate relations should be
reliable only up to ${\cal O}(1)$.

\noindent
{\it 9. A consistency check} may be performed by comparing the left-hand
sides of the last two equations.

\bigskip
\leftline{\bf D. $B \to D \bar D$ and $B \to \eta_c P$ decays}
\bigskip

Here we must discuss the relations implied by Tables~\ref{tabvii-dd}
and~\ref{tabviii-dd} separately, since a single factor of $\lambda$ no
longer relates the two. Although $\that'/\that \simeq \chat'/\chat \simeq
\ehat'/\ehat \simeq - \lambda$ (and similarly for the corresponding
SU(3)-breaking terms), the ratio $\phat'/\phat$ is expected to be only of
order $\lambda$, but not to the same accuracy.

By now our methods should have become clear to the reader, but we enumerate
the consequences of the tables explicitly for the sake of completeness.
\smallskip

\noindent
{\it 1. Numerous isospin relations} may be written. These consist of the
amplitude relations
\beq
A(B^+ \to D_s^+ D^0) = A(B^0 \to D_s^+ D^-)~~~,
\eeq
\beq
A(B^+ \to \eta_c K^+) = A(B^0 \to \eta_c K^0)~~~,
\eeq
\beq
A(B_s \to D^+ D^-) = - A(B_s \to D^0 \bar D^0)~~~
\eeq
\beq
A(B^+ \to \eta_c \pi^+) = \s A(B^0 \to \eta_c \pi^0)~~~,
\eeq
and the triangle relation
\beq \label{eqn:bddtri}
A(B^+ \to D^+ \bar D^0) = A(B^0 \to D^+ D^-) + A(B^0 \to D^0 \bar D^0)~~~.
\eeq
The consequences of the $I = 0$ nature of the $\bar b \to \bar c c \bar s$
transition for $B \to K J/\psi$ decays were pointed out some time ago
\cite{LS}.
\smallskip

\noindent
{\it 2. The effects of color-suppressed and exchange-type SU(3)-breaking
amplitudes} can be extracted {}from the data:
\beq
\lambda^2 \Gamma(B^+ \to \eta_c K^+) / \Gamma(B^+ \to \eta_c \pi^+)
= 1 + 2 {\rm~Re}(\chat_1/\chat)~~~,
\eeq
\beq
\Gamma(B_s \to \eta_c \bar K^0) / \Gamma(B^+ \to \eta_c \pi^+)
= 1 + 2 {\rm~Re}(\chat_2/\chat)~~~,
\eeq
\beq
\Gamma(B^0 \to D_s^+ D_s^-) / \Gamma(B^0 \to D^0 \bar D^0)
= 1 + 2 {\rm~Re}(\ehat_2/\ehat)~~~,
\eeq
\beq
\lambda^2 \Gamma(B_s \to D^+ D^-) / \Gamma(B^0 \to D^0 \bar D^0)
= 1 + 2 {\rm~Re}(\ehat_1/\ehat)~~~.
\eeq

\noindent
{\it 3. The effects of the combination $\that_2' + \phat_2'$} can be
extracted {}from the ratio
\beq
\Gamma(B_s \to D^+ D_s^-) / \Gamma(B^+ \to D^+ \bar D^0)
= 1 + 2 {\rm~Re}[(\that_2' + \phat_2')/(\that' + \phat')]~~~.
\eeq

If we now ignore exchange-type diagrams, we find several more relations:
\smallskip

\noindent
{\it 4. Several amplitudes vanish.}  Thus,
\beq
A(B_s \to D^+ D^-) = A(B_s \to D^0 \bar D^0)
= A(B^0 \to D_s^+ D_s^-) = A(B^0 \to D^0 \bar D^0) = 0~~~.
\eeq
As one consequence, the triangle relation (\ref{eqn:bddtri}) becomes
an amplitude equality:
\beq
A(B^+ \to D^+ \bar D^0) = A(B^0 \to D^+ D^-)~~~.
\eeq

\noindent
{\it 5. The effects of the combination $\that_2 + \phat_2$} can be
extracted {}from the ratio
\beq
\Gamma(B_s \to D_s^+ D_s^-)/\Gamma(B^+ \to D_s^+ \bar D^0)
= 1 + 2 {\rm~Re}[(\that_2 + \phat_2)/(\that + \phat)]~~~,
\eeq
where we recall that we are working only to first order in SU(3) breaking.

The tree contributions $\that$ and $\that'$ always occur in combination with
the corresponding penguin terms $\phat$ and $\phat'$. A number of additional
consequences would follow if we were to assume that $\phat'/\phat \simeq
\that'/\that$, or that the penguin terms (which must produce a $c \bar c$ pair)
are negligible. In the latter case one could determine the ratio
$(f_{D_s}/f_D)^2$ [see Eq.~(\ref{fdsfd})] by comparing $\Gamma(B^+ \to D^+ \bar
D^0)$ with $\lambda^2 \Gamma(B^+ \to D_s^+ \bar D^0)$. Other rate ratios which
can be used to obtain $(f_{D_s}/f_D)^2$ are $\Gamma(B^0 \to D^+D^-)/\lambda^2
\Gamma(B^0\to D_s^+ D^-)$ and $\Gamma(B_s \to D^+D_s^-)/\lambda^2 \Gamma(B_s\to
D_s^+ D_s^-)$.
\bigskip

\centerline{\bf V. SU(3) BREAKING AND THE EXTRACTION OF CKM PHASES}
\bigskip

In Refs.~\cite{PRL,PLB} we presented a number of ways to extract CKM phases,
strong phases, and the sizes of individual diagrams from $B\to PP$ decays. All
these analyses made use of unbroken SU(3) symmetry (as well as the neglect of
$E$, $A$ and $PA$ diagrams) to relate $B\to\pi\pi$, $B\to\pi K$ and $B \to K
\bar K$ decays. In this section we discuss the implications of SU(3)-breaking
effects for such analyses. (Note that electroweak penguins \cite{DH,EWP}, which
we neglect here, may be of equal, or greater, importance than SU(3)-breaking
effects. If such contributions are large, they may well invalidate the analyses
of Refs.~\cite{PRL,PLB}. However, if they are small, then SU(3) breaking is the
important factor, which is why it is useful to consider it separately, as we do
here.)

Ref.~\cite{PRL} makes use of the SU(3) triangle relation of
Eq.~(\ref{eqn:su3tri}), rewritten below for convenience:
\[
A(B^+ \to \pi^+ K^0) + \s A(B^+ \to \pi^0 K^+) = \lambda \s A(B^+ \to \pi^+
\pi^0)~~~.
\]
If $A$-type diagrams are neglected, these three amplitudes have the following
graphical decomposition:
\beqn
A(B^+\to \pi^+\pi^0) & = & -{1\over\sqrt{2}}(T+C)~~~, \nonumber \\
A(B^+\to \pi^+ K^0) & = & P'~~~, \\
A(B^+\to \pi^0 K^+) & = & -{1\over\sqrt{2}}(T'+C'+P')~~~. \nonumber
\eeqn
Now consider the triangle formed from the three CP-conjugate processes:
\beq
A(B^- \to \pi^- {\bar K}^0) + \s A(B^- \to \pi^0 K^-) =
\lambda \s A(B^- \to \pi^- \pi^0)~~~.
\eeq
The $P'$ amplitude is dominated by the CKM matrix elements $V^*_{tb}
V_{ts}$, whose phase is $\pi$. Thus, this amplitude is common to both
triangles:
\beq
A(B^+\to \pi^+ K^0) = A(B^-\to \pi^- {\bar K}^0).
\eeq
The weak phase of the $T+C$ amplitude is $\gamma$. Thus we have
\beq
|A(B^+\to \pi^+\pi^0)| = |A(B^-\to \pi^-\pi^0)|.
\eeq
The third amplitude, $T'+C'+P'$, has two contributions ($(T'+C')$ and $P'$)
with different weak and strong phases. Hence there can be CP violation in
the decay $B^\pm \to \pi^0 K^\pm$.

\begin{figure}
\centerline{\psfig{figure=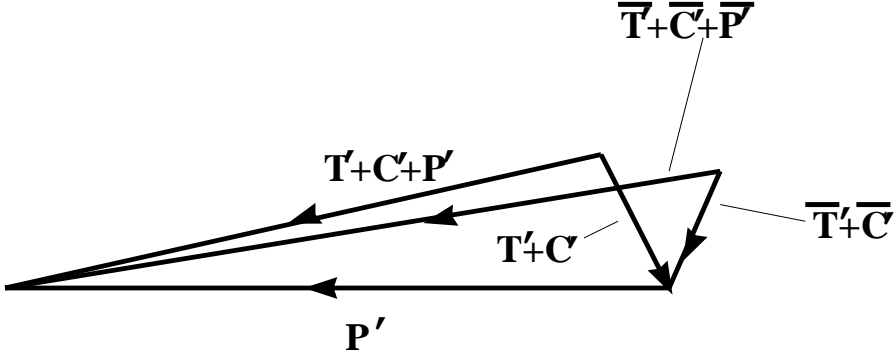,width=14cm}}
\caption{Triangle relating amplitudes $T'+C'+P'= -\protect \sqrt{2} A(B^+ \to
\pi^0 K^+)$, $P'= A(B^+ \to \pi^+ K^0)$, and $T'+C'= \lambda (f_K/f_\pi)
A(B^+ \to  \pi^+ \pi^0)$, as well as the corresponding charge-conjugate
processes (denoted by bars over symbols for amplitudes). The angle between
$T'+C'$ and $\overline{T'}+\overline{C'}$ is $2\gamma$.}
\label{PRLfig}
\end{figure}

When one compares the triangle to the CP-conjugate triangle, the angle between
the amplitudes $\lambda \s A(B^+\to \pi^+\pi^0)$ and $\lambda \s A(B^-\to
\pi^-\pi^0)$ is just $2\gamma$ (see Fig.~\ref{PRLfig}). There is a twofold
ambiguity corresonding to the interchanging of $\gamma$ and $\delta_{TC'} -
\delta_{P'}$, where $\delta_{TC'}$ and $\delta_{P'}$ are the strong phases of
the  $(T'+C')$ and $P'$ amplitudes, respectively.

How does this analysis hold up when we consider SU(3) breaking? From Tables
\ref{tabi-pp} and \ref{tabii-pp}, the decomposition of the amplitudes in
terms of SU(3)-invariant and SU(3)-breaking contributions is
\beqn
A(B^+\to \pi^+\pi^0) & = & -{1\over\sqrt{2}}(T+C)~~~, \nonumber \\
A(B^+\to \pi^+ K^0) & = & P'+P_1'~~~, \\
A(B^+\to \pi^0 K^+) & = & -{1\over\sqrt{2}}(T'+C'+P'+T_1'+C_1'+P_1')~~~.
\nonumber
\eeqn
In other words, the third side of the triangle is really
$-(T'+C'+T_1'+C_1')$, whereas above we assumed it was $-\lambda (T+C) =
-(T'+C')$. The error incurred is simply
\beq
1 + {T_1' + C_1' \over T' + C'} \simeq 1 + {T_1' \over T'} ~~~,
\eeq
where, on the right-hand side, we have neglected the $C'$ and $C_1'$ terms
as being subdominant (see Sec.~III B). (This approximation is at the same
level as the neglect of $A$-type diagrams.) However, from
Eq.~(\ref{fkfpi}), assuming factorization, we have
\beq
1 + {T_1' \over T'} = {f_K\over f_\pi} e^{i\delta_{SU(3)}}~~~,
\eeq
where we have included a possible additional strong phase. This simply
reflects the fact that, in the presence of SU(3) breaking, the $T$ and $T'$
amplitudes no longer have the same strong phase: $\delta_{T'} = \delta_T +
\delta_{SU(3)}$. Therefore, taking into account SU(3)-breaking effects,
Eq.~(\ref{eqn:su3tri}) should read
\beq
A(B^+ \to \pi^+ K^0) + \s A(B^+ \to \pi^0 K^+) = {f_K\over f_\pi}
e^{i\delta_{SU(3)}} \lambda \s A(B^+ \to \pi^+ \pi^0)~~~.\nonumber
\eeq
This does not change things substantively. $\delta_{TC'}$ and $\delta_{P'}$ are
now the strong phases of $(T'+C'+T_1'+C_1')$ and $P'+P_1'$, respectively. Apart
from this, the analysis of Ref.~\cite{PRL} still holds, as long as the factor
$f_K/f_\pi$ is included. The weak phase $\gamma$ can be obtained, up to a
twofold ambiguity which interchanges it and the strong phase $\delta_{TC'} -
\delta_{P'}$.

\begin{figure}
\centerline{\psfig{figure=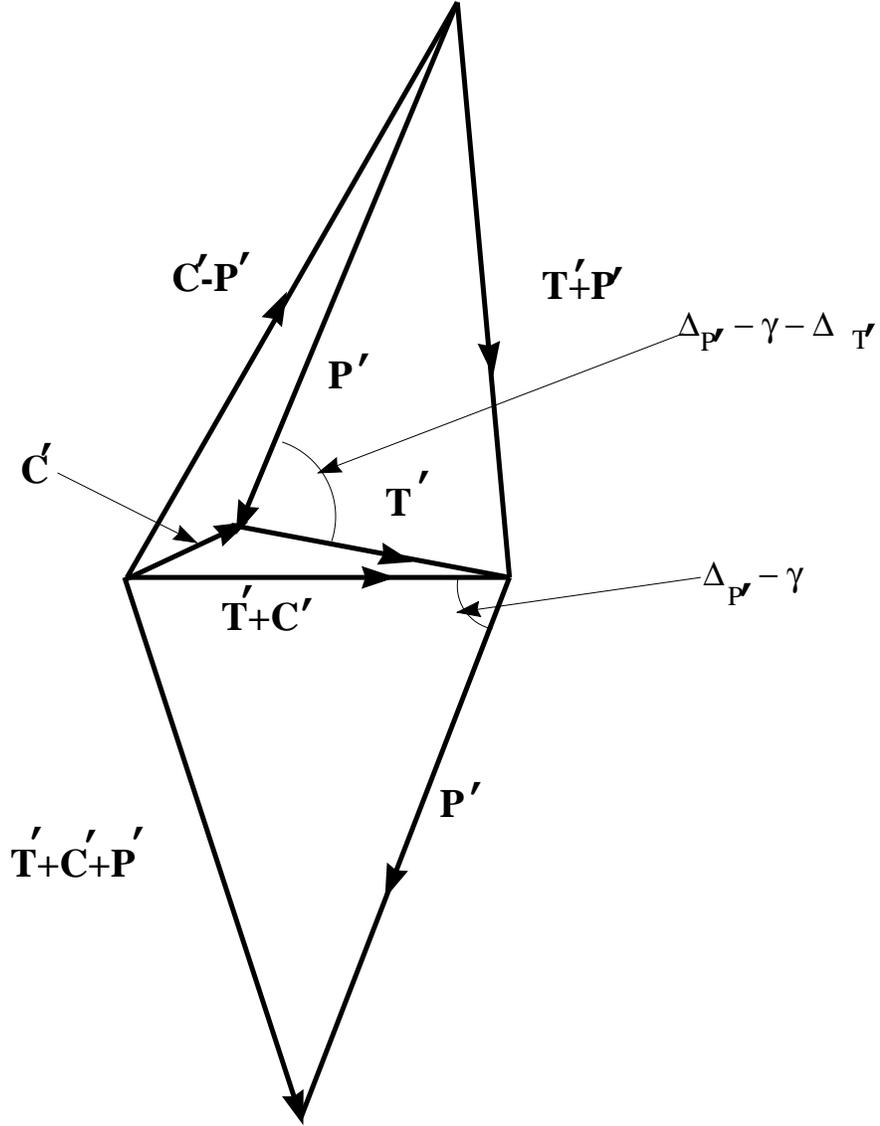,height=16cm}}
\caption{Amplitude triangles based on Eqs.~(17) and (18) permitting the
extraction of strong phases and the weak phase $\gamma$ in the SU(3)
symmetry limit and with linear SU(3) symmetry breaking.}
\label{PLBfiga}
\end{figure}

Ref.~\cite{PLB} describes two types of analyses. The first is essentially
an extension of the analysis described above, except that it also allows
one to extract the strong phases and sizes of the individual $T'$, $C'$ and
$P'$ diagrams. It makes use of the isospin $\pi K$ quadrangle
[Eq.~(\ref{quadrangle})], rewritten below for convenience:
\[
A(B^+ \to \pi^+ K^0) + \s A(B^+ \to \pi^0 K^+)
= A(B^0 \to \pi^- K^+) + \s A(B^0 \to \pi^0 K^0)~~~.
\]
The key point is that, within SU(3) symmetry, one diagonal of the $\pi K$
quadrangle is related to the amplitude for $B^+ \to \pi^+\pi^0$
[Eq.~(\ref{eqn:su3tri})]: $(T'+C') = \lambda (T+C)$. Thus, by measuring the
four $\pi K$ rates, as well as the rate for $B^+ \to \pi^+\pi^0$, one can
construct the diagram in Fig.~\ref{PLBfiga}. This allows the extraction of
$|P'|$, $|T'|$, $|C'|$, $\Delta_{C'}$ and $\Delta_{T'}$, as well as the
quantity $\Delta_{P'} - \Delta_{T'} - \gamma$, where $\Delta_i \equiv
\delta_i - \delta_{TC'}$. If one also measures the CP-conjugate processes,
$\gamma$ can be disentangled from $\Delta_{P'} - \Delta_{T'}$, as in
\cite{PRL}. (It should be pointed that there is some uncertainty in the
determination of $|C'|$ and $\Delta_{C'}$. Although the amplitude $A'$ is
negligible compared to $P'$ and $T'$, it is not so small when compared to
$C'$ -- we estimate $|A'/C'| \sim \lambda$. Thus, the precision in our
determinations of $|C'|$ and $\Delta_{C'}$ is limited by the neglect of
$A'$.)

If one considers SU(3) corrections, Fig.~\ref{PLBfiga} still holds, except that
(i) $P'$, $T'$ and $C'$ now include their SU(3) corrections $P_1'$, $T_1'$ and
$C_1'$, respectively, and (ii) the diagonal is no longer $(T'+C')$, which is
directly related to $(T+C)$, but rather $(T'+C'+T_1'+C_1')$. However, we showed
above how to relate this SU(3)-corrected diagonal to $(T+C)$: up to small
corrections, $|T'+C'+T_1'+C_1'| = (f_K/f_\pi) \lambda |T+C|$. Thus, the
analysis still holds, except that the strong phases that are measured include
SU(3)-breaking effects. (To be precise, the SU(3) correction $C_1'$ should be
neglected everywhere. It is expected to be of the same order as $E$ and
$A$-type diagrams, which have been ignored. This means that, just as in the
SU(3)-invariant case, the determination of $|C'|$ is accurate to only about
25\%, and $\Delta_{C'}$ is similarly affected.)

The second analysis of Ref.~\cite{PLB} is a bit more complicated. If one
assumes unbroken SU(3) symmetry, one has two triangles with a common base
[see Eqs.~(\ref{eqn:ispintri}), (\ref{quadrangle}) and (\ref{eqn:su3tri})]:
\beqn
\lambda \s A(B^+ \to \pi^+ \pi^0) & = & \lambda A(B^0 \to \pi^+ \pi^-) +
\lambda \s A(B^0 \to \pi^0 \pi^0)~~~, \nonumber \\
\lambda \s A(B^+ \to \pi^+ \pi^0) & = & A(B^0 \to \pi^- K^+) +
\s A(B^0 \to \pi^0 K^0)~~~.
\label{eqn:PLBtrib}
\eeqn
In terms of diagrams, these two triangles can be written
\beqn
\lambda (T+C) & = & \lambda (T+P) + \lambda (C-P) ~~~, \nonumber \\
\lambda (T+C) & = & (T'+P') + (C'-P') ~~~.
\eeqn

\begin{figure}
\centerline{\psfig{figure=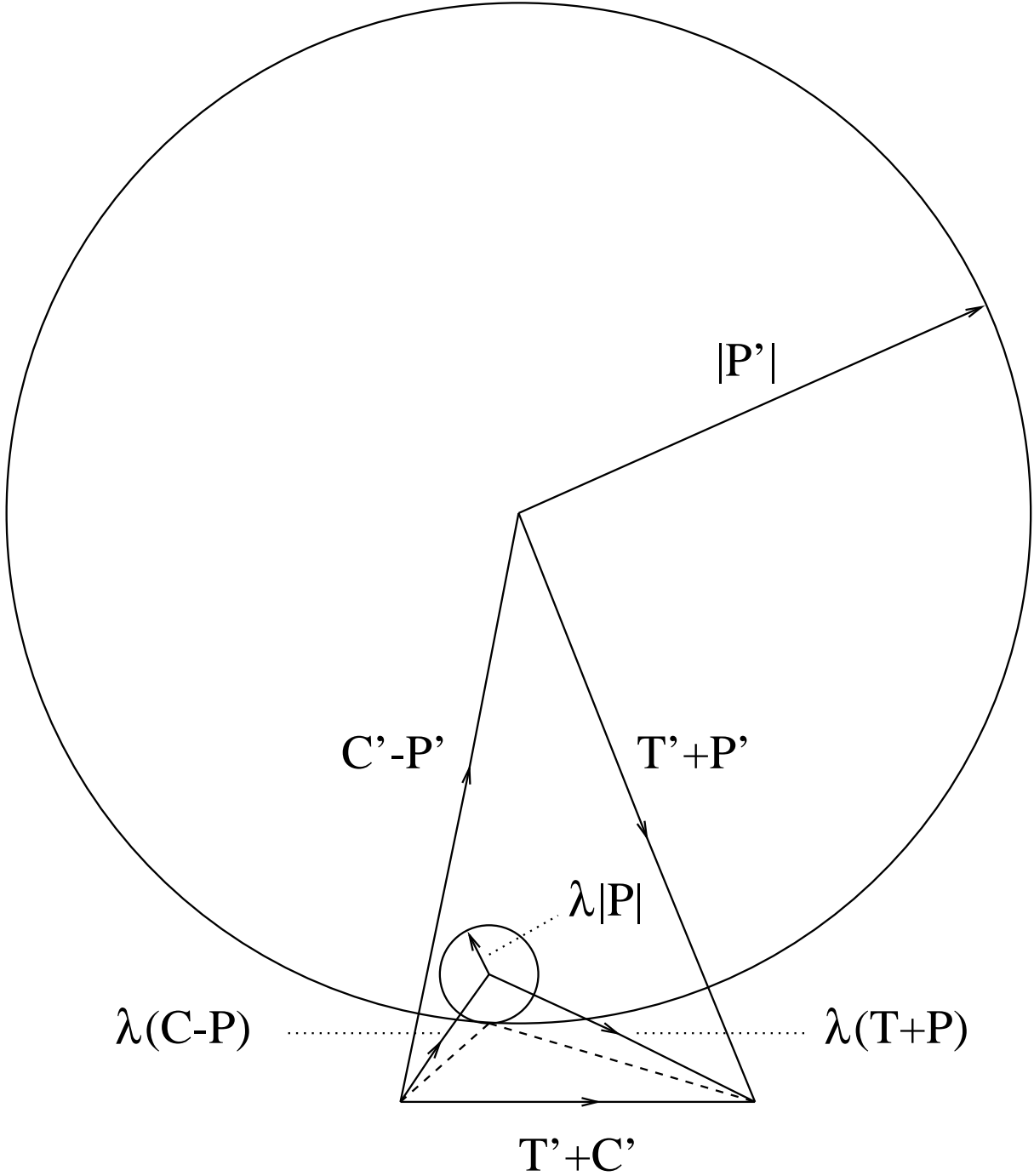,width=16cm,angle=90}}
\caption{Amplitude triangle based on Eqs.~(\protect\ref{eqn:PLBtrib})
permitting the extraction of strong phase shift differences and the weak phases
$\beta, \gamma$ in the SU(3) symmetry limit. With linearly broken SU(3), only
the extraction of $\gamma$ and certain strong phases is possible (see text).}
\label{PLBfigb}
\end{figure}

By measuring the rates for $B^+ \to \pi^+ K^0$ and $B^0 \to K^0 {\bar K}^0$,
one can obtain the magnitudes of $P'$ and $P$, respectively. With these 7 rate
measurements, one can construct the diagram of Fig.~\ref{PLBfigb}, in which the
apex of the subtriangle $T'+C' = (T'+C')$ is determined, up to a twofold
ambiguity, from the intersection of the two circles. The key point here is that
this fixes the {\it orientation} of the vectors $P$ and $P'$. Thus we can
obtain their phases, relative to the $(T'+C')$ amplitude (the horizontal line).
These relative phases are $\Delta_P + \alpha$ and $\Delta_{P'} - \gamma$,
respectively (we have assumed that the weak phase of the $P$ diagram is given
approximately by Arg$(V_{tb}^* V_{td}) = -\beta$ \cite{burasfleischer}, and we
have used $\alpha=\pi-\beta-\gamma$). However, within SU(3) symmetry, $\Delta_P
= \Delta_{P'}$, so that one can combine these two phase measurements to obtain
the weak CKM phase $\beta$. In addition, one can also obtain the strong phases
and magnitudes of the various diagrams. If one also measures the rates for the
CP-conjugate processes, it is possible to obtain $\gamma$, $\alpha$ and
$\Delta_P$ separately. (Note that the precision with which the magnitude and
phase of $C' = \lambda C$ can be determined is limited as in the first
construction by the neglect of $A$- and $PA$-type diagrams.)

Unfortunately, in the presence of SU(3)-breaking effects, this analysis does
not stand up as well as the previous two. We will be able to extract $\gamma$
and certain strong phases in a way independent of the previous constructions,
but we will not be able to obtain the other CKM phases.

There are two places where SU(3)-breaking effects effects are important. First,
$P$ and $P'$ get different SU(3) corrections: the amplitude $P'$ in the decay
$B^+ \to \pi^+ K^0$ gets an SU(3) correction $P_1'$, while the amplitude $P$ in
the decay $B^0 \to K^0 {\bar K}^0$ has a $P_3$ correction. Thus, the equality
$\Delta_P = \Delta_{P'}$ is likely to be broken, so that the CKM angle $\beta$
cannot be extracted as described above. Furthermore, the $P_3$ correction to
$P$ is not present in the isospin triangle [Eq.~(\ref{eqn:ispintri})]. This
means that there is some uncertainty as to the position of the apex of the
subtriangle. Thus, the orientation of the $P+P_3$ vector is poorly determined
-- even if we somehow knew that $\Delta_P = \Delta_{P'}$, we still could not
obtain $\beta$ precisely. (Note that the orientation of $P'+P_1'$ can still be
fairly accurately obtained -- since $|P|\ll|P'|$, a small correction to $P$ has
very little effect on the orientation of $P'+P_1'$ as determined from the
intersection of the $P$ and $P'$ circles in Fig.~\ref{PLBfigb}.)

Second, there are really two subtriangles: $T + C = (T+C)$ and $T' + C' =
(T'+C')$. Assuming a perfect SU(3) symmetry, these subtriangles are congruent,
and simply scale by $\lambda$. However, this is no longer true in the presence
of SU(3) breaking. We know how to take certain SU(3)-breaking effects into
account. For example, assuming factorization, $T'$ and $T$ are related by
$(f_K/f_\pi) \exp(i\delta_{\rm SU(3)})$, as are $(T'+C')$ and $(T+C)$ to a good
approximation (the error is at the level of $\sim\lambda^2$ relative to the
dominant $T$ and $T'$ diagrams.) However, $C'$ and $C$ are not so clearly
related. {\it A priori}, we do not know the relation between these two
amplitudes. In this case, the small difference between $(T'+C')$ and $(T+C)$
can have a significant effect. Since the $C$ diagram is smaller than the $T$
diagram by a factor of $\lambda$, the small error one makes in relating
$(T'+C')$ to $(T+C)$ can be a large error in the determination of the
magnitudes and phases of $C$ and $C'$ (in addition to the error incurred by
neglecting $A$- and $PA$-type diagrams). This in turn leads to a further
uncertainty in the position of the apex of the subtriangle.

On the other hand, since $|T|$ and $|T'|$ are much larger than $|C|$ and
$|C'|$, respectively, the small uncertainty in the position of the apex of the
subtriangle has little effect on the determination of $|T'|$ and $\Delta_{T'}$.
Thus, the quantity $\Delta_{P'} - \Delta_{T'} - \gamma$ can be extracted in the
same way as in the first analysis of Ref.~\cite{PLB}. If one measures the
CP-conjugate processes, one can similarly disentangle $\gamma$ and $\Delta_{P'}
- \Delta_{T'}$. One of the advantages of this method over the previous one was
that the weak phase $\beta$ could be obtained. In the presence of SU(3)
breaking this is no longer the case. Furthermore, the determinations of $|C'|$
and $\Delta_{C'}$ remain imprecise. However even in the presence of SU(3)
breaking, this method can still be used to {\it independently} determine
$\gamma$ and some of the strong phases.
\bigskip

\centerline{\bf VI. FINAL-STATE INTERACTIONS}
\bigskip

\leftline{\bf A. $B \to \pi \bar D$ decays}
\bigskip

The decays $B^+ \to \pi^+ \bar D^0$, $B^0 \to \pi^+ D^-$, and $B^0 \to
\pi^0 \bar D^0$ involve one amplitude leading to a final state with $I =
1/2$ and one amplitude leading to a final state with $I = 3/2$.
Specifically, the weak Hamiltonian for the transition $\bar b \to \bar c u
\bar d$ transforms as $I = I_3 = 1$, permitting the following decomposition
of the amplitudes in terms of the $\pi \bar D$ isospins:
$$
A(B^+ \to \pi^+ \bar D^0) = A_{3/2}~~~,
$$
$$
A(B^0 \to \pi^+ D^-) = (2/3)A_{1/2} + (1/3)A_{3/2}~~~,
$$
\beq
\s A(B^0 \to \pi^0 \bar D^0) = -(2/3)A_{1/2} + (2/3)A_{3/2}~~~.
\eeq

These amplitudes clearly satisfy a triangle relation, as already written in
(\ref{eqn:pid}). Since a single CKM element dominates the decays, a
non-zero area for this triangle would signify differences in final-state
phases between the $I = 1/2$ and $I = 3/2$ amplitudes. This circumstance
has been used by H. Yamamoto \cite{HY} to place upper limits on such phase
differences, not only in the decays $B \to \pi \bar D$, but also in $B \to
\pi \bar D^*$ and $B \to \rho \bar D$. A similar method has already been
used in the decays $D \to \pi \bar K$ and related processes to conclude
that there are important final-state phase differences between the $I =
1/2$ and $I = 3/2$ $\pi \bar K$ and $\pi \bar K^*$ states \cite{charmfs}.

\begin{figure}
\centerline{\psfig{figure=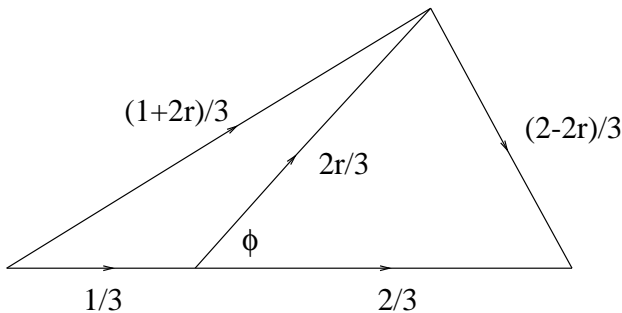,height=4.0cm,angle=90}}
\caption{Amplitude triangle for determining the phase of $r \equiv A_{1/2}/
A_{3/2}$ in $B \to \pi \bar D$ decays.}
\label{figviii-tri}
\end{figure}

We illustrate in Fig.~\ref{figviii-tri} an amplitude triangle for $B \to \pi
\bar D$ decays, where we define $r \equiv A_{1/2}/A_{3/2}$. The base of the
triangle has unit length, while the two other sides have lengths
\beq
\frac{A(B^0 \to \pi^+ D^-)}{A(B^+ \to \pi^+ \bar D^0)} = \frac{1 + 2r}{3}
\eeq
and
\beq
\frac{\sqrt{2} A(B^0 \to \pi^0 \bar D^0)}{A(B^+ \to \pi^+ \bar D^0)}
= \frac{2 - 2r}{3}~~~.
\eeq
A line drawn from a point 1/3 of the way along the base to the apex then
has the phase $\phi \equiv {\rm Arg}(r)$.

\bigskip

\newpage
\leftline{\bf B. $B \to K \bar D$ decays}
\bigskip

A single CKM matrix element, governing the transition $\bar b \to \bar c u
\bar s$, also dominates the decays $B^+ \to K^+ \bar D^0$, $B^0 \to K^+
D^-$, and $B^0 \to K^0 \bar D^0$. The weak Hamiltonian transforms as $I =
I_3 = 1/2$. The decay amplitudes may be decomposed into contributions with
final-state isospins $I = 0$ and $I = 1$:
$$
A(B^+ \to K^+ \bar D^0) = A_1'~~~,
$$
$$
A(B^0 \to K^+ D^-) = (1/2)A_1' + (1/2)A_0'~~~,
$$
\beq
A(B^0 \to K^0 \bar D^0) = (1/2)A_1' - (1/2)A_0'~~~.
\eeq
Thus, they will satisfy a triangular relation (\ref{eqn:kd}). If the
triangle has non-zero area, final-state interactions are important.
Similar results apply, for example, to $B \to K^* \bar D$ and $B \to K \bar
D^*$ decays.
\bigskip

\leftline{\bf C. SU(3) relations between $B \to \pi \bar D$ and $B \to K
\bar D$ decays}
\bigskip

The results of Tables~\ref{tabiii-pdbar} and~\ref{tabiv-pdbar} imply
relations among the amplitudes for $B \to \pi \bar D$ and $B \to K \bar D$
decays. In the absence of SU(3) breaking and exchange diagram ($E$)
contributions, we would expect $A_1' = \lambda A_{3/2}$ and $A_0' = (1/3)
\lambda (4 A_{1/2} - A_{3/2})$. By comparing the expressions for the
respective decays in terms of amplitudes $T$ and $C$ or $T'$ and $C'$, we
see that if the triangles (\ref{eqn:pid}) and (\ref{eqn:kd}) have {\it
different shapes}, one must conclude that (i) SU(3) is broken, (ii)
exchange contributions are important, or (iii) both.
\bigskip

\leftline{\bf D. Other tests for final-state interactions}
\bigskip

The decays $B \to \pi \bar D$ and $B \to K \bar D$ offer the best hope of
providing clean tests for final-state interactions with reasonable decay
rates and triangles whose sides are all expected to be non-vanishing.
However, two additional amplitude triangles and one amplitude quadrangle
may be of use in testing for final-state interactions. These are the
relations (\ref{eqn:bpdtri}), (\ref{eqn:bddtri}), and (\ref{eqn:bpdquad})
involving the decays $B \to K D$, $B \to D \bar D$, and $B \to \pi D$,
respectively.

Since the decay $B^+ \to K^0 D^+$ is expected to proceed purely through an
annihilation diagram (see Table~\ref{tabv-pd}), the triangle containing
this amplitude should have one very short side. It may be very difficult
to tell that such a triangle has non-zero area. Similarly, the decay $B^0
\to D^0 \bar D^0$ should proceed purely via an exchange diagram
(Table~\ref{tabviii-dd}), so its triangle may have a short side. The
amplitude quadrangle (\ref{eqn:bpdquad}) applies to the decays $B \to \pi
D$ whose amplitudes are of order $\lambda^4$, and hence not likely to be
detected soon. One could tell if such a quadrangle had non-zero area by
constructing its sides as the square roots of observed rates and checking
that no two or three sides added up to any other two or one side.
\bigskip

\leftline{\bf E. Comments on rescattering effects}
\bigskip

In Ref.~\cite{BPP,PRL,PLB} the neglect of $E$, $A$, and $PA$ contributions in
comparison with $T$, $C$, and $P$ contributions was noted explicitly to be
equivalent to the assumption that certain rescattering effects are unimportant.
For example, a final state which can be reached through the annihilation
diagram can also be reached through a tree diagram followed by a rescattering.
Several tests of this hypothesis were proposed in Ref.~\cite{BPP,PRL,PLB}.  It
is no surprise that this assumption leads to relations between final-state
phases in different decay channels. Indeed one such phase relation was noted to
exist between $B\to \pi\pi$ and $B \to \pi K$ \cite{where}.

It was subsequently pointed out \cite{LWC} that a relation among final-state
phases was implicit in assuming that the decay the decay $B^+ \to \pi^+ K^0$ is
pure penguin (here we have neglected the annihilation diagram).  This is indeed
so.  The point raised in Ref.~\cite{LWC} is that the $I = 3/2$ combination
\beq
A(\pi^+ K^0) + \sqrt{2} A(\pi^0 K^+) = - (T' + C')
\label{eq-athreehalfs}
\eeq
and the $I = 1/2$ tree contribution to the combination
\beq
[2 A(\pi^+ K^0) - \sqrt{2} A(\pi^0 K^+)]_{\rm tree} = + (T' + C')
\label{eq-ahalf}
\eeq
should have the same strong final-state phases if their sum is to vanish. In
the graphical description of Ref.~\cite{BPP,PRL,PLB}, this is automatically the
case, since the amplitude in Eq.~(\ref{eq-athreehalfs}) and the tree
contribution to the combination in Eq.~(\ref{eq-ahalf}) are both proportional
to $T' + C'$. Thus, the equivalence of the strong final-state phases is a
direct consequence of our assumption that the annihilation diagrams are
negligible.

We stress that our general treatment of linearly broken flavor SU(3) in
two-body $B$ decays does not forbid final-state interaction phases. Although
OZI-forbidden scattering from one $q \bar q$ pair to another is not permitted
at the quark level by our decomposition \cite{Lipkin}, each of the hadronic
decay amplitudes, $T$, $C$, $P$, etc.\ may carry a nonzero CP-conserving phase.
For example, part of the phase of $P$ can be calculated perturbatively
\cite{BSS}.
\bigskip

\newpage
\centerline{\bf VII. EXPERIMENTAL DATA AND PROSPECTS}
\bigskip

In this section we give a snapshot of the present status of data. We include
results which are anticipated soon with the events in hand. We then discuss
briefly the improvements which would be needed to test various sectors of the
theory. Our treatment of $B_s$ decays is rather sketchy since it is premature
to assess experimental possibilities until more final states have been
reconstructed.
\bigskip

\leftline{\bf A. Decays to two light mesons}
\bigskip

Here we concentrate mainly on expected hierarchies of the dominant amplitudes
$T,T'$, $C,C'$, and $P,P'$, and the potential for confirming them. We have
already mentioned in Ref.~\cite{BPP} the (rather demanding) levels of
statistics required to test for the presence of the diagrams $E,E'$, $A,A'$,
and $PA,PA'$.

Some combination of the decays $B^0 \to \pi^- \pi^+$ and $B^0 \to \pi^- K^+$
has been observed \cite{Kpisep}, with a combined branching ratio of about $2
\times 10^{-5}$. Equal mixtures of the two modes are likely, though a decisive
separation awaits better particle identification. It then appears
\cite{BPP,SilWo} that the amplitude $T$ dominates the $B^0 \to \pi^- \pi^+$
decay, while $P'$ dominates $B^0 \to \pi^- K^+$ (see Tables \ref{tabi-pp} and
\ref{tabii-pp}), as we estimated in Sec.~III B.

Other $B \to PP$ decays which should be visible at branching ratios of 1/2 to
$1 \times 10^{-5}$ (depending on whether they involve a neutral or charged
pion) are $B^+ \to \pi^+ \pi^0$, $B_s \to \pi^+ K^-$, and all the remaining
processes in Table \ref{tabii-pp}. For example, if the $C$ amplitude is small,
one expects
\beq \label{eqn:pirel}
\Gamma(B^+ \to \pi^+ \pi^0) \approx \Gamma(B^0 \to \pi^- \pi^+)/2~~~.
\eeq
If the $P'$ amplitude is the only one present in $B \to \pi K$, one expects
$$
\Gamma(B^+ \to \pi^+ K^0) \approx 2 \Gamma(B^0\to \pi^0 K^0)
$$
\beq \label{eqn:krel}
\approx \Gamma(B^0 \to \pi^- K^+) \approx 2 \Gamma(B^+\to \pi^0 K^+)~~~.
\eeq
Present upper limits on branching ratios (at the 90\% confidence level) for
such processes include \cite{lims}
$B(B^+ \to \pi^+ \pi^0) < 2.3 \times 10^{-5}$,
$B(B^0 \to \pi^0 K^0) < 6.3 \times 10^{-5}$,
$B(B^+ \to \pi^0 K^+) < 3.2 \times 10^{-5}$, and
$B(B^+ \to \pi^+ K^0) < 6.8 \times 10^{-5}$,
with no information available for $B_s$ decays. Updated results for some
of these modes are forthcoming \cite{Wuert}.

The $\Delta S = 0$ processes in Table \ref{tabi-pp} containing only
color-suppressed and/or penguin contributions, such as $B^0 \to \pi^0 \pi^0$,
should be characterized by branching ratios of order $10^{-6}$ or smaller (see
also \cite{BSW}). In this class of processes, Ref.~\cite{lims} quotes only the
limit $B(B^0 \to \pi^0 \pi^0) < 1.0 \times 10^{-5}$. Thus, one must wait for an
improvement of about a factor of ten in present data before expecting to see
such processes consistently. At this level, one will be able to construct a
meaningful triangle based on the three distinct decay rates for $B \to \pi
\pi$, and one should expect deviations from the relation (\ref{eqn:pirel}).

A factor of ten increase in data will also permit the observation of rate
differences in the various $B \to \pi K$ channels, as a consequence of
interference of the term $T'$ in Table \ref{tabii-pp} with the dominant $P'$
term. If $C'$ is sufficiently small in comparison with $T'$, these rate
differences should violate the middle equality in (\ref{eqn:krel}) while
preserving the other two equalities:
\beq \label{eqn:krela}
\Gamma(B^+ \to \pi^+ K^0) \approx 2 \Gamma(B^0\to \pi^0 K^0)
\eeq
and
\beq \label{eqn:krelb}
\Gamma(B^0 \to \pi^- K^+) \approx 2 \Gamma(B^+\to \pi^0 K^+)~~~.
\eeq
Electroweak penguin contributions \cite{DH,EWP} could disturb these
relations, making them of particular interest for early testing.

In $B$ decays to one light vector meson and one light pseudoscalar, there are
hints of signals in several $B \to \pi K^*$ channels \cite{lims}. However, as
noted (e.g.) in \cite{DZ} and \cite{BPP}, the SU(3) analysis in these channels
is more involved, so we have not undertaken a general treatment of
SU(3)-breaking effects. Some partial results on the role of electroweak
penguins have been obtained \cite{EWP}.
\bigskip

\leftline{\bf B. $B \to P \bar D$ decays}
\bigskip

We begin by discussing the ${\cal O}(\lambda^2)$ processes in Table
\ref{tabiii-pdbar}.

The color-favored decays of nonstrange $B$ mesons, involving the amplitude
$\bar T$, have been seen at branching ratio levels of 1/4 to nearly 2 \%
\cite{Alam}, in the $\pi \bar D$, $\pi \bar D^*$, $\rho \bar D$, and $\rho \bar
D^*$ channels. Typical upper limits on the color-suppressed $B^0$ decays to
these channels are an order of magnitude lower. As noted in \cite{HY}, one can
already construct meaningful amplitude triangles for several of these channels,
placing upper limits on the relative phase shifts between $I = 1/2$ and $I =
3/2$ channels which are typically tens of degrees.

What level of data would be required to see effects of the $\bar E$
contribution? The amplitude for such a process is expected to be only a few
percent of the dominant $\bar T$ contribution. The equality of $\Gamma(B^0 \to
\pi^+ D^-) \sim |\bar T + \bar E|^2$
with $\Gamma(B_s \to \pi^+ D_s^-) \sim |\bar
T + \bar T_2|^2$ is more likely to be
upset by the SU(3)-breaking term $\bar T_2$
than by the term $\bar E$. So far one candidate for $B_s \to \pi^+ D_s^-$ has
been seen \cite{ALEPHBs}.

In order to see the effect of $\bar E$ alone, one would have to detect the
decay $B^0 \to K^+ D_s^-$ (or a related process involving one or more vector
mesons). The present limits \cite{Alex} of
$B(B^0 \to K^+ D_s^-) < 2.3 \times 10^{-4}$,
$B(B^0 \to K^+ D_s^{*-}) < 1.7 \times 10^{-4}$,
$B(B^0 \to K^{*+} D_s^-) < 9.7 \times 10^{-4}$,
$B(B^0 \to K^{*+} D_s^{*-}) < 1.1 \times 10^{-3}$
are not adequate to detect the presence of the $\bar E$ contribution at the
predicted level. The present upper limits on $|\bar E/ (\bar T + \bar E)| <
1/\sqrt{12}$ from $B(B^0\to K^+ D_s^{*-})/B(B^0 \to \pi^+ D^{*-})$ and on
$|\bar E/ (\bar T + \bar C)| < 1/\sqrt{20}$ from $B(B^0\to K^+ D_s^-)/B(B^+
\to \pi^+ \bar D^0)$ must be improved considerably for an observation of
decay modes dominated by $E$ and $A$ amplitudes, if these terms are indeed
suppressed by $f_B/m_B \sim \lambda^2$.

None of the strangeness-changing $B \to P \bar D$ decays listed in Table
\ref{tabiv-pdbar} has been reported yet. The observation of the decay $B^+
\to K^+ \bar D^0$ probably offers the best prospects. If SU(3) breaking
can be accounted for by the ratio $f_K/f_\pi$, as one expects to be true
for the dominant $\bar T$ contribution, one expects
\beq
\frac{\Gamma(B^+ \to K^+ \bar D^0)}{\Gamma(B^+ \to \pi^+ \bar D^0)}
= \frac{|f_K~V_{us}|^2}{|f_\pi~V_{ud}|^2} \approx 0.075~~~,
\eeq
while this ratio would be only about 0.051 in the absence of SU(3)
breaking.

Since about 300 $B^+ \to \pi^+ \bar D^0$ events have already been reported by
the CLEO II \cn~\cite{Alam}, there should be nearly two dozen events of $B^+
\to K^+ \bar D^0$ in the same sample. An observed sample of some hundred $B^+
\to K^+ \bar D^0$ events would be able to test conclusively for the SU(3)
breaking mentioned above.

In the absence of appreciable $\bar E$ contributions, one should expect
\beq
\frac{\Gamma(B^0 \to K^+ D^-)}{\Gamma(B^0 \to \pi^+ D^-)}
= \frac{|f_K~V_{us}|^2}{|f_\pi~V_{ud}|^2} \approx 0.075~~~
\eeq
as well. About 80 events of $B^0 \to \pi^+ D^-$ have been reported by CLEO
II so far \cite{Alam}.
\medskip

\leftline{\bf C. $B \to P D$ decays}
\bigskip

Here one is dealing with amplitudes which, though nominally of order
$\lambda^3$ (Table \ref{tabv-pd}) or $\lambda^4$ (Table \ref{tabvi-pd}),
may be further suppressed by the smallness of $V_{ub}$ and the effects of
form factors. Nonetheless, it is important to detect modes such as the
color-suppressed decay $B^+ \to K^+ D^0$ if the program of Ref.~\cite{MG}
for determining the weak phase $\gamma$ is to be implemented.

The process with the best prospect of being seen first is the decay $B^0
\to \pi^- D_s^+$, for which there exists only the upper limit of $2.7
\times 10^{-4}$ on the branching ratio \cite{Alex}. A crude estimate based
on factorization in which we neglect form factor differences and
color-suppressed diagrams would predict
\beq
\frac{\Gamma(B^0 \to \pi^- D_s^+)}{\Gamma(B^0 \to \pi^- \pi^+)}
\approx \frac{f_{D_s}^2}{f_\pi^2} \approx 5~~~,
\eeq
where we have taken $f_{D_s} \approx 300$ MeV. Thus, we expect a branching
ratio for $B^0 \to \pi^- D_s^+$ of several parts in $10^5$. The decay
should begin to show up with several times the present data sample. At
precisely half the rate of $B^0 \to \pi^- D_s^+$, (as a consequence of
isospin), one should see the decay $B^+ \to \pi^0 D_s^+$.

Observation of the color-suppressed $B \to K D$ decays will require a
further increase of about tenfold in the data. At this level one may test
SU(3) by comparing the processes involving $\tt + \tt_1$ or $\tt + \tt_1 +
\tt_2$ in Table \ref{tabv-pd} with those involving $\lambda \tt$ or
$\lambda(\tt + \tt_2)$ in Table \ref{tabvi-pd}.
\medskip

\leftline{\bf D. $B \to D \bar D$ decays}
\bigskip

Decays such as $B \to D_s^+ \bar D$ (see Table \ref{tabvii-dd}) (and the
corresponding processes involving one or two vector mesons) have been
observed with branching ratios of 1 -- 2 \% \cite{CLEODs}. Somewhat over
100 events have been observed in the sum of all channels. Isospin
invariance predicts pairwise equalities for charged and neutral $B$ decay
modes.

The color-suppressed decays $B \to J/\psi K^{(*)}$ have been observed with
branching ratios which are about an order of magnitude smaller than those
of $B \to D^{*}_s\bar D^{(*)}$. This provides information about the ratio
$|\chat/\that|$ which is somewhat larger than $\lambda$. Similar branching
ratios, of about $10^{-3}$, are expected for $B \to \eta_c K^{(*)}$ which
should soon be observed through the hadronic decay modes of the $\eta_c$.

The decays $B^+ \to D^+ \bar D^0$ and $B^0 \to D^+ D^-$ (see Table
\ref{tabviii-dd}) should occur at several percent of the rates for $B^+ \to
D_s^+ \bar D^0$ and $B^0 \to D_s^+ D^-$, with precise ratios dictated by
ratios of heavy meson decay constants if a factorization hypothesis is
adequate to describe these decays and if penguin amplitudes are negligible.

The presence of $\ehat$ contributions would be most cleanly illustrated by
observing decays of the form $B_s \to D \bar D$. With $f_B/m_B \approx
5\%$, we estimate the corresponding branching ratio to be at most a few
parts in $10^5$. Present fragmentary information on $B_s$ meson production
does not allow us to estimate the size of the data sample that would permit
such a test.
\bigskip

\leftline{\bf E. Overall prospects}
\bigskip

The present sample of $B$ decays is based in large part on the 2 million
nonstrange $B \bar B$ pairs collected so far by CLEO,
with impressive reconstructions of some decay modes
(including those of $B_s$) by groups at LEP and by the CDF Collaboration
at Fermilab. A foreseen upgrade
of the luminosity of CESR to ${\cal L} = 10^{33}~{\rm cm}^{-2} s^{-1}$
should provide 10 million such pairs in a year ($10^7$ s) of operation.
Asymmetric $B$ factories at SLAC and KEK should provide comparable (or
eventually larger) samples. Nonetheless, it seems hard to escape the
conclusion that many of the tests proposed here will require larger data
sets than can be achieved at electron-positron colliders. The ability of
hadron colliders to produce large numbers of $B$ mesons is unquestioned; it
remains to be seen whether a large enough fraction of these can be
detected.

\bigskip
\centerline{\bf VIII. CONCLUSIONS}
\bigskip

We have discussed prospects for experimental tests of several aspects of
two-body hadronic $B$ decays, including SU(3)-breaking, the neglect of
certain SU(3) amplitudes corresponding to disfavored graphs, and the
elucidation of strong final-state-interaction phase differences. While
decays to pairs of light pseudoscalar mesons typically involve more than
one product of elements of the Cabibbo-Kobayashi-Maskawa (CKM) matrix,
decays in which one or two of the final quarks are charmed typically have a
simpler CKM structure. Consequently, the effects of interest to us can be
more readily isolated.

We have discussed a staged set of measurements, starting with the present
sample of nonstrange $B$ decays (dominated by CLEO II data) and progressing
through the multiplication of this sample by successive factors. Results
which may be testable in the near future include the following:

1) We have presented a diagrammatic description of the various
SU(3)-breaking effects. Assuming factorization for $T$-type diagrams, one
SU(3)-breaking diagram corresponds to the ratio of decay constants. Using
this description, we expect that $\Gamma(B^+ \to K^+ \bar
D^{(*)0})/\Gamma(B^+ \to \pi^+ \bar D^{(*)0}) = |f_K V_{us}|^2 / |f_\pi
V_{ud}|^2 \approx 0.075$, while this ratio would be only about 0.051 in the
absence of SU(3) breaking. Similar comments apply to the ratio $\Gamma(B^0
\to K^+ D^{(*)-})/\Gamma(B^+ \to \pi^+ D^{(*)-})$.

2) The study of $B \to D \bar D$ decays (Tables 7 and 8) can provide
information on the ratio $f_D/f_{D_s}$ if factorization is assumed:
$\Gamma(B^+ \to D^+ \bar D^0) / \Gamma(B^+ \to D_s^+ \bar D^0)$,
$\Gamma(B^0 \to D^+D^-)/ \Gamma(B^0\to D_s^+ D^-)$ and
$\Gamma(B_s \to D^+D_s^-)/ \Gamma(B_s\to D_s^+ D_s^-)$ are all expected to
equal $|f_D V_{cd}|^2 / |f_{D_s} V_{cs}|^2$. This same ratio of CKM matrix
elements and decay constants can also be obtained from
$\Gamma(B^0 \to \pi^- D^+)/ \Gamma(B^0 \to \pi^- D_s^+)$, but this is
likely to be less useful experimentally, since a small [${\cal
O}(\lambda^4)$] amplitude is involved.

3) Other SU(3)-breaking effects, associated with form factors and quark
pair creation, can also be isolated by ratios of rate measurements. The
list of such measurements is very long, so we refer the reader to Sec.~IV
for a complete discussion.

4) A search for decays such as $B^0 \to K^{(*)+} D_s^{(*)-}$ at an order of
magnitude better sensitivity than present levels will start to shed light
on the presence or absence of weak $B$ meson decays involving the light
spectator quark. Other processes of order $\lambda^2$ in the amplitude
which are of this type are the decays $B_s \to D^+ D^-$ and $B_s \to D^0
\bar D^0$ (Table 7).

5) The processes in 4) are all of the ``exchange'' type. In order to look
for purely ``annihilation'' amplitudes one must turn to the process $B^+
\to K^0 D^+$ (Table 5), of order $\lambda^3$. This process is involved in
an isospin triangle relation together with the decays $B^+ \to K^+ D^0$ and
$B^0 \to K^0 D^0$. Unequal rates for these last two decays also would be
evidence for the annihilation contribution.

6) Other ${\cal O}(\lambda^3)$ processes of the purely ``exchange''
variety include $B_s \to \pi^+ D^-$ and $B_s \to \pi^0 {\bar D}^0$ (Table
4), $B_s \to \pi^- D^+$ and $B_s \to \pi^0 D^0$  (Table 5), and $B^0 \to
D^0 \bar D^0$ and $B^0 \to D_s^+ D_s^-$ (Table 8).  These should also be
suppressed.

7) Some SU(3) relations which should hold even in the presence of SU(3)
breaking (but whose validity depends on the neglect of exchange and
annihilation contributions) have been obtained, including the amplitude
relation $A(B^+ \to K^+ \bar K^0) = A(B^0 \to K^0 \bar K^0)$ (see Sec.~IV).

8) We find that the program for obtaining the weak phase $\gamma$, in several
independent ways, from $B \to PP$ decays described in Refs.~\cite{PRL} and
\cite{PLB} is not substantially affected by a more careful consideration of
SU(3) breaking (see Sec.~V).  Some strong phase information can also still be
extracted. On the other hand, the determination of $\beta$ proposed in
Ref.~\cite{PLB} is much more vulnerable to such effects. The role of
electroweak penguins in such determinations has been discussed in a separate
paper \cite{EWP}.

9) Triangle relations involving the decays $B \to \pi \bar D$ \cite{HY} and
$B \to K \bar D$ (and related states involving vector mesons) will provide
useful information on strong final-state phase shift differences, since
these decays are dominated by a single CKM matrix element.

10) A hierarchy of contributions to various decays has been discussed (Sec.~III
B), whereby one can estimate the expected rates for rare processes without
reference to specific models. Rates of color-suppressed decays are expected to
be intermediate between rates of color-favored processes and processes
dominated by ``annihilation" or ``exchange" amplitudes.

To sum up, a rich set of questions may be addressed by measurements of
rates for two-body $B$ decays, from the present levels which include
branching ratios of more than a percent down to levels of $10^{-7}$ or
lower. Eventually, one will want to detect large numbers of $B_s$ decays in
order to fully implement this program.

\newpage 
\centerline{\bf ACKNOWLEDGMENTS}
\bigskip

We thank J. Cline, A. Dighe, I. Dunietz, G. Eilam, A. Grant, K. Lingel, H.
Lipkin, R. Mendel, S. Stone, L. Wolfenstein, and M. Worah for fruitful
discussions. J. Rosner wishes to acknowledge the hospitality of the Fermilab
theory group and the Cornell Laboratory for Nuclear Studies during parts of
this investigation. M. Gronau, O. Hern\'andez and D. London are grateful for
the hospitality of the University of Chicago, where part of this work was done.
This work was supported in part by the United States -- Israel Binational
Science Foundation under Research Grant Agreement 90-00483/3, by the
German-Israeli Foundation for Scientific Research and Development, by the Fund
for Promotion of Research at the Technion, by the NSERC of Canada and les Fonds
FCAR du Qu\'ebec, and by the United States Department of Energy under Contract
No. DE FG02 90ER40560.

\def \ajp#1#2#3{Am.~J.~Phys.~{\bf#1}, #2 (#3)}
\def \apny#1#2#3{Ann.~Phys.~(N.Y.) {\bf#1}, #2 (#3)}
\def \app#1#2#3{Acta Phys.~Polonica {\bf#1}, #2 (#3)}
\def \arnps#1#2#3{Ann.~Rev.~Nucl.~Part.~Sci.~{\bf#1}, #2 (#3)}
\def \cmp#1#2#3{Commun.~Math.~Phys.~{\bf#1}, #2 (#3)}
\def \cmts#1#2#3{Comments on Nucl.~Part.~Phys.~{\bf#1}, #2 (#3)}
\def \corn93{{\it Lepton and Photon Interactions:  XVI International
Symposium, Ithaca, NY August 1993}, AIP Conference Proceedings No.~302,
ed.~by P. Drell and D. Rubin (AIP, New York, 1994)}
\def \cp89{{\it CP Violation,} edited by C. Jarlskog (World Scientific,
Singapore, 1989)}
\def \dpff{{\it The Fermilab Meeting -- DPF 92} (7th Meeting of the
American Physical Society Division of Particles and Fields), 10--14
November 1992, ed. by C. H. Albright \ite~(World Scientific, Singapore,
1993)}
\def \dpf94{DPF 94 Meeting, Albuquerque, NM, Aug.~2--6, 1994}
\def \efi{Enrico Fermi Institute Report No. EFI}
\def \el#1#2#3{Europhys.~Lett.~{\bf#1}, #2 (#3)}
\def \f79{{\it Proceedings of the 1979 International Symposium on Lepton
and Photon Interactions at High Energies,} Fermilab, August 23-29, 1979,
ed.~by T. B. W. Kirk and H. D. I. Abarbanel (Fermi National Accelerator
Laboratory, Batavia, IL, 1979}
\def \hb87{{\it Proceeding of the 1987 International Symposium on Lepton
and Photon Interactions at High Energies,} Hamburg, 1987, ed.~by W. Bartel
and R. R\"uckl (Nucl. Phys. B, Proc. Suppl., vol. 3) (North-Holland,
Amsterdam, 1988)}
\def \ib{{\it ibid.}~}
\def \ibj#1#2#3{~{\bf#1}, #2 (#3)}
\def \ichep72{{\it Proceedings of the XVI International Conference on High
Energy Physics}, Chicago and Batavia, Illinois, Sept. 6--13, 1972,
edited by J. D. Jackson, A. Roberts, and R. Donaldson (Fermilab, Batavia,
IL, 1972)}
\def \ijmpa#1#2#3{Int.~J.~Mod.~Phys.~A {\bf#1}, #2 (#3)}
\def \ite{{\it et al.}}
\def \jmp#1#2#3{J.~Math.~Phys.~{\bf#1}, #2 (#3)}
\def \jpg#1#2#3{J.~Phys.~G {\bf#1}, #2 (#3)}
\def \lkl87{{\it Selected Topics in Electroweak Interactions} (Proceedings
of the Second Lake Louise Institute on New Frontiers in Particle Physics,
15--21 February, 1987), edited by J. M. Cameron \ite~(World Scientific,
Singapore, 1987)}
\def \ky85{{\it Proceedings of the International Symposium on Lepton and
Photon Interactions at High Energy,} Kyoto, Aug.~19-24, 1985, edited by M.
Konuma and K. Takahashi (Kyoto Univ., Kyoto, 1985)}
\def \mpla#1#2#3{Mod.~Phys.~Lett.~A {\bf#1}, #2 (#3)}
\def \nc#1#2#3{Nuovo Cim.~{\bf#1}, #2 (#3)}
\def \np#1#2#3{Nucl.~Phys.~{\bf#1}, #2 (#3)}
\def \pisma#1#2#3#4{Pis'ma Zh.~Eksp.~Teor.~Fiz.~{\bf#1}, #2 (#3) [JETP
Lett. {\bf#1}, #4 (#3)]}
\def \pl#1#2#3{Phys.~Lett.~{\bf#1}, #2 (#3)}
\def \plb#1#2#3{Phys.~Lett.~B {\bf#1}, #2 (#3)}
\def \pr#1#2#3{Phys.~Rev.~{\bf#1}, #2 (#3)}
\def \pra#1#2#3{Phys.~Rev.~A {\bf#1}, #2 (#3)}
\def \prd#1#2#3{Phys.~Rev.~D {\bf#1}, #2 (#3)}
\def \prl#1#2#3{Phys.~Rev.~Lett.~{\bf#1}, #2 (#3)}
\def \prp#1#2#3{Phys.~Rep.~{\bf#1}, #2 (#3)}
\def \ptp#1#2#3{Prog.~Theor.~Phys.~{\bf#1}, #2 (#3)}
\def \rmp#1#2#3{Rev.~Mod.~Phys.~{\bf#1}, #2 (#3)}
\def \rp#1{~~~~~\ldots\ldots{\rm rp~}{#1}~~~~~}
\def \si90{25th International Conference on High Energy Physics, Singapore,
Aug. 2-8, 1990}
\def \slc87{{\it Proceedings of the Salt Lake City Meeting} (Division of
Particles and Fields, American Physical Society, Salt Lake City, Utah,
1987), ed.~by C. DeTar and J. S. Ball (World Scientific, Singapore, 1987)}
\def \slac89{{\it Proceedings of the XIVth International Symposium on
Lepton and Photon Interactions,} Stanford, California, 1989, edited by M.
Riordan (World Scientific, Singapore, 1990)}
\def \smass82{{\it Proceedings of the 1982 DPF Summer Study on Elementary
Particle Physics and Future Facilities}, Snowmass, Colorado, edited by R.
Donaldson, R. Gustafson, and F. Paige (World Scientific, Singapore, 1982)}
\def \smass90{{\it Research Directions for the Decade} (Proceedings of the
1990 Summer Study on High Energy Physics, June 25 -- July 13, Snowmass,
Colorado), edited by E. L. Berger (World Scientific, Singapore, 1992)}
\def \stone{{\it B Decays}, edited by S. Stone (World Scientific,
Singapore, 1994)}
\def \tasi90{{\it Testing the Standard Model} (Proceedings of the 1990
Theoretical Advanced Study Institute in Elementary Particle Physics,
Boulder, Colorado, 3--27 June, 1990), edited by M. Cveti\v{c} and P.
Langacker (World Scientific, Singapore, 1991)}
\def \yaf#1#2#3#4{Yad.~Fiz.~{\bf#1}, #2 (#3) [Sov.~J.~Nucl.~Phys.~{\bf #1},
#4 (#3)]}
\def \zhetf#1#2#3#4#5#6{Zh.~Eksp.~Teor.~Fiz.~{\bf #1}, #2 (#3) [Sov.~Phys.
- JETP {\bf #4}, #5 (#6)]}
\def \zpc#1#2#3{Zeit.~Phys.~C {\bf#1}, #2 (#3)}

\end{document}